\documentclass[%
showpacs,preprintnumbers,
 amsmath,amssymb,
 aps,
prc,
 lengthcheck%
]{revtex4-1}

\usepackage{graphicx}
\usepackage{dcolumn}
\usepackage{bm}
\usepackage{hyperref}
\usepackage{comment}
\usepackage{braket}
\usepackage{mathtools}

\def\nuc#1#2{\relax\ifmmode{}^{#1}{\protect\text{#2}}\else${}^{#1}$#2\fi}

\newcommand{\be}{\begin{eqnarray}}
\newcommand{\ee}{\end{eqnarray}}
\newcommand{\tj}[6]{
\begin{pmatrix}
  #1 & #2 & #3 \\
  #4 & #5 & #6 
 \end{pmatrix}}

\newcommand{\sj}[6]{ \begin{Bmatrix}
  #1 & #2 & #3 \\
  #4 & #5 & #6 
 \end{Bmatrix}}

\newcommand{\nj}[9]{ \begin{Bmatrix}
  #1 & #2 & #3 \\
  #4 & #5 & #6 \\
  #7 & #8 & #9
 \end{Bmatrix}}

\usepackage{graphicx}
\graphicspath{ {/home/mario/paperTExc/figures/} }
\begin{document}

\title{Interplay of projectile breakup and target excitation in reactions induced by weakly-bound nuclei}


\author{M. G\'omez-Ramos}
\email{mgomez40@us.es}
\author{A.~M.\ Moro}
\email{moro@us.es}
\affiliation{Departamento de FAMN, Facultad de F\'{\i}sica, Universidad de Sevilla,%
Apdo.~1065, E-41080 Sevilla, Spain}

\vspace{1cm}

\date{\today}


\begin{abstract}
\begin{description}
\item[Background]  Reactions involving weakly-bound nuclei require formalisms able to deal with continuum states. The majority of these formalisms struggle to treat collective excitations of the systems involved. For CDCC, extensions to include target excitation have been developed but have only been applied to a small number of cases.
\item[Purpose]  In this work, we reexamine the extension of the CDCC formalism to include target excitation and apply it to a variety of reactions to study the effect of breakup on inelastic cross sections.
\item[Methods] We use a transformed oscillator basis to discretize the continuum of the projectiles in the different reactions and use the extended CDCC method developed in this work to solve the resulting coupled differential equations. A new code has been developed to perform the calculations.
\item[Results] Reactions $^{58}{\rm Ni}(d,d)^{58}{\rm Ni}^*$, $^{24}{\rm Mg}(d,d)^{24}{\rm Mg}^*$, $^{144}{\rm Sm}(^{6}{\rm Li},^{6}{\rm Li})^{144}{\rm Sm}^*$ and $^{9}{\rm Be}(^{6}{\rm Li},^{6}{\rm Li})^{9}{\rm Be}^*$ are studied. Satisfactory agreement is found between experimental data and extended CDCC calculations.
\item[Conclusions] The studied CDCC method has proven to be an accurate tool to describe target excitation in reactions with weakly-bound nuclei. Moderate effects of breakup on inelastic observables are found for the reactions studied. Cross section magnitudes are not modified much, but angular distributions present smoothing when opposed to calculations without breakup.
\end{description}
\end{abstract}


\maketitle

\section{\label{intro} Introduction}

Few-body models have been very successful in describing nuclear reactions involving weakly-bound nuclei, where breakup probabilities are high and continuum states of these nuclei influence heavily other relevant reaction channels such as elastic scattering. Many models able to deal with these positive-energy states have been developed, such as Continuum-Discretized Coupled Channels (CDCC) \citep{PhysRevC.9.2210CDCC,Austern1987125CDCC}, the adiabatic approximation \citep{PhysRevC.61.047301,PhysRevC.57.3225}, Faddeev/AGS equations \citep{faddeev60,Alt} and several semi-classical approximations \citep{PhysRevC.50.2104, Esbensen199637, PhysRevC.50.R1276, PhysRevC.64.024601, PhysRevC.70.064605, GarcíaCamacho2006118}.

In most of these models, in particular CDCC, the weakly-bound nucleus is considered to be composed of two subsystems, the valence particle and the core, which may be dissociated during the reaction in a breakup process. These subsystems are usually considered to be inert at the energies of interest. This is a good approximation for reactions involving deuterons, which were the origin of many of these models, but it is more questionable for more complex systems. For this reason, extensions to include the collective excitations of the core subsystem have been developed both for CDCC \citep{summers, deDiego14} and Faddeev/AGS equations \citep{Del13}. These extensions were able to give proper descriptions of reactions involving weakly-bound nuclei with deformed cores such as \nuc{11}{Be} \citep{Mor12b} and \nuc{19}{C} \citep{PhysRevC.94.021602}.

In general, these models also consider that the target is an inert system without internal degrees of freedom relevant for the reaction. Possible excitations of the target are assumed to be effectively included in the fragment-target optical potentials, rather than explicitly treated. This assumption makes these models unsuitable to describe excitation of the target or more generally any process in which both breakup of one system and collective excitation of the other may take place concurrently.

However, there exist a variety of measurements in which weakly-bound nuclei collide with nuclei with collective degrees of freedom which are excited during the reaction, either due to a big deformation and an associated rotational spectrum or to the existence of low-energy vibrational levels \citep{Kiss19761, Woodard201217}. These experiments require a consistent description of both breakup and collective excitation of the target if reliable information is to be obtained from them.

Extensions to include target excitation in CDCC were already developed in the 80s by the Kyushu-Pittsburg groups \citep{Yahiro01041986}, and have received some recent attention \citep{PierreChau2015}, but in general have been restricted to deuteron scattering, only to its $s$-wave component, and mutual excitation has been ignored in order to make the calculations computationally feasible. Therefore, we find it timely to reexamine the corresponding formalism and apply it to more general reactions using the full formalism without introducing further approximations. In this respect, it must be remarked that the extension of the Faddeev/AGS equations which allowed for the inclusion of collective excitations of the core subsystem also permits the inclusion of target excitation thanks to the symmetric treatment of target and projectile which is employed in Faddeev equations \citep{Del13}. In this work we reexamine the extension of the CDCC method to include target excitation. The outline of this work is the following: in section \ref{sec:theory} the formalism used for the extension is presented, while calculations for different low- to medium-energy reactions are shown in section \ref{sec:calc}. Finally, the summary and conclusions are presented in section \ref{sec:summary}.

\section{\label{sec:theory}  Scattering framework}  

In this section we derive the expression for the coupling potentials which allow to treat breakup of the projectile and excitation of the collective states of the target on an equal footing. Although these potentials were already derived by the Kyushu-Pittsburg group \citep{Yahiro01041986}, we find it suitable to present them again here, in order to make this work more self-consistent and to avoid difficulties that may arise due to the different coupling schemes used here and in \citep{Yahiro01041986}. We have chosen the $j-j$ coupling to apply the formalism to nuclei heavier than the deuteron, in contrast to \citep{Yahiro01041986}, who employed the $L-S$ coupling scheme, which is more commonly used to describe the different components of the deuteron. 

In our framework, the projectile is modelled as a two body system composed of a core  ($c$) and a valence particle ($v$), $\vec{r}$ being their relative coordinate. The target is assumed to have some internal degrees of freedom $\xi$, which may be excited due to its interaction with the projectile. The effective Hamiltonian of the projectile-target system is of the form:
\begin{equation}
\label{eqSchr}
H(\vec{R},\vec{r},\xi)=T(\vec{R})+h_p(\vec{r})+h_t(\xi)+V_{pt}(\vec{r},\vec{R},\xi),
\end{equation}
where $\vec{R}$ is the relative coordinate between projectile and target, $T$ is the kinetic energy operator of the projectile-target system and $h_p$ is the internal Hamiltonian of the projectile:
 \begin{equation}
h_p(\vec{r})=T(\vec{r})+V_{vc}(\vec{r}),
\end{equation}
which depends on the relative coordinate between valence and core systems. We must remark that in this formalism $V_{vc}$ may depend on other internal coordinates of core or valence particle, resulting in eigenstates of the projectile which are a linear combination of single-particle components. However, it is assumed here that these internal coordinates will not be modified by the interaction with the core, so their influence is reduced to the structure of the projectile and will not affect the dynamics of the reaction. Thus, they are ignored in our derivation. Nevertheless, we note that the following derivation is fully applicable to multicomponent projectile wavefunctions, but components with different states of the core will not be coupled in this formalism.

In Eq.~(\ref{eqSchr}), $h_t$ is the internal Hamiltonian of the target, which only depends on collective degrees of freedom, whose identities will depend on the model chosen to describe the target. Finally, the interaction potential $V_{pt}$ is divided in the interaction between the valence particle and the target $V_{vt}$, and between the core and the target $V_{ct}$. Both interactions depend on the relative coordinate between each subsystem and the target and the internal degrees of freedom of the target, namely:
\begin{eqnarray}
V_{pt}(\vec{r},\vec{R},\xi)=V_{vt}(\vec{r_v},\xi)+V_{ct}(\vec{r_c},\xi) \\
\vec{r_v}=\vec{R}-\gamma_v\vec{r} \hspace{1cm}
\vec{r_c}=\vec{R}-\gamma_c\vec{r} \\
 \gamma_v=\dfrac{m_c}{m_c+m_v} \hspace{1cm} \gamma_c=-\dfrac{m_v}{m_c+m_v} \label{coords}
\end{eqnarray}
The projectile wavefunctions are the eigenfunctions of $h_p$, and a process of discretization of the continuum is employed to treat breakup states \citep{Mor12b,Austern1987125CDCC,particlemotion,deDiego14}. The index $i$ is used to denote the states resulting from this discretization procedure. The $j-j$ coupling system is chosen to express the projectile wavefunctions:
\begin{equation}
\Phi^i_{J_p}(\vec{r})=\sum_{l,j,I}\dfrac{\varphi^{i}_{J_pljI}(r)}{r}
\left[\mathcal{Y}_{lsj} \otimes \chi_I \right]_{J_p},
\label{projwf}
\end{equation}
where $\varphi^{i}_{J_pljI}$ is the radial internal wavefunction of the projectile for a definite $l$ (orbital angular momentum of the valence-core system); $\vec{j}=\vec{l}+ \vec{s}$, where $s$ is the spin of the valence particle; and $I$, the spin of the core. $\mathcal{Y}_{lsj}$ is the spin spherical harmonic and $\chi_I$ is the wavefunction of the core (core and valence states are assumed to be completely determined by their spins).

The target wavefunctions ($\Phi^n_{J_t}(\xi)$) are the eigenstates of $h_t$, depend on $\xi$, have an angular momentum $J_t$ and are completely defined by the quantum number(s) $n$.

The scattering wavefunctions are chosen as in \citep{Mor12b, deDiego14, fresco}
\begin{align}
 \Psi_{J_T,M_T}(\vec{R},\vec{r},\xi)&=  \sum_{\beta} \chi_\beta^{J_T} (R)\nonumber\\ \times& \left\lbrace \left[ Y_L(\hat{R}) \otimes \Phi^i_{J_p}(\vec{r}) \right]_J \otimes \Phi^n_{J_t}(\xi) \right\rbrace_{J_T,M_T},
\end{align}
where $\beta$ denotes all the quantum numbers necessary to define the channel $\beta=\lbrace L,i,J_p,J,J_t,n \rbrace$. From now on we will be using a notation similar to that employed in \citep{deDiego14, summers}:
\small
\begin{equation}
\left\langle\hat{R},\vec{r},\xi|\beta,J_TM_T\right\rangle=\left\lbrace \left[ Y_L(\hat{R}) \otimes \Phi^i_{J_p}(\vec{r}) \right]_J \otimes \Phi^n_{J_t}(\xi) \right\rbrace_{J_T,M_T}.
\end{equation}
\normalsize
The most important physical ingredients for our calculations are the coupling potentials:
\small
\begin{equation}
\label{coupPot}
U^{J_T}_{\beta,\beta'} (R)=\left\langle \beta,J_TM_T|V_{vt}(\vec{R},\vec{r},\xi)+V_{ct}(\vec{R},\vec{r},\xi)|\beta',J_TM_T\right\rangle.
\end{equation}
\normalsize
It is assumed that both $V_{vt}$ and $V_{ct}$ can be expanded in multipoles of $\vec{r}$ such as:
\begin{equation}
V_{vt}(\vec{r}_{v})=\sqrt{4\pi}\sum_{Qq}V_{Qq}(r_{v},\xi)Y_{Qq}(\hat{r}_{v}),
\end{equation} 
and likewise for $V_{ct}$. In some common models for nuclear excitations, such as the rotational or vibrational models \cite{bohrmottelson}, it is possible to factorize the potential in radial and structure parts, leading to:
\begin{equation}
V_{vt}(\vec{r}_{v})=\sqrt{4\pi}\sum_{Qq}V^Q_{vt}(r_{v})\mathcal{T}^*_{Qq} (\xi)Y_{Qq}(\hat{r}_{v}),
\label{multexpV}
\end{equation} 
where the radial part is $V^Q_{vt}(r_{v})$ and $\mathcal{T}^*_{Qq} (\xi)$ is an operator with the same tensorial character as $Y_{Qq}$, appropriate to the structure model used to describe the target.


The coupling potentials (\ref{coupPot}) can be expressed as:
\begin{widetext}
\begin{equation}
\begin{split}
\label{eqU}
&U^{J_T}_{\beta,\beta'} (R)=\braket{\beta\|V_{vt}+V_{ct}\|\beta'}=\sum_{\Lambda,\Lambda',Q}(-1)^{Q+J_T+J'+J_t}\hat{J}\hat{J'}\sj{J'}{J_t'}{J_T}{J_t}{J}{Q}\nj{L'}{J_p'}{J'}{\Lambda}{\Lambda'}{Q}{L}{J_p}{J}
\dfrac{1}{\sqrt{4\pi}}\hat{L}\hat{\Lambda} \langle L0\Lambda0|L'0\rangle F^{\Lambda,\Lambda',Q}_{\beta\beta'} (R),
\end{split}
\end{equation}
\end{widetext}
where 6j, 9j symbols and Clebsch-Gordan coefficients appear as usual and  $\hat{A}=\sqrt{2A+1}$. Primed quantities refer to the initial channel and unprimed ones to the final one. $\Lambda, \Lambda'$ and $Q$ are the orbital, projectile and target angular momentum transferred, respectively, and the radial form factors $F^{\Lambda,\Lambda',Q}_{\beta\beta'} (R)$ do not depend on the orbital angular momentum $L$. The expression for these form factors is the following:
\begin{widetext}
\begin{align}
\label{eqF}
F^{\Lambda,\Lambda',Q}_{\beta\beta'} (R)&=(-1)^{J_p'+\Lambda+J_t'-J_t}
\sqrt{4\pi}\hat{J_p}\hat{J_p'}\hat{l}\hat{l}'\hat{j}\hat{j'}\hat{\Lambda}\hat{\Lambda}'^2
\hat{Q}^2\sum_{\alpha,\alpha',K,\lambda} (-1)^{j'+j+l+l'+s+I} \hat{K}^2\begin{pmatrix} Q \\
 \lambda\end{pmatrix}\tj{l'}{l}{\Lambda'}{0}{0}{0} \tj{\lambda}{K}{\Lambda}{0}{0}{0}\nonumber  \\\times&\tj{Q-\lambda}{K}{\Lambda'}{0}{0}{0} \sj{\Lambda'}{\Lambda}{Q}{\lambda}{Q-\lambda}{K}\sj{j'}{j}{\Lambda'}{l}{l'}{s}
\sj{J_p'}{J_p}{\Lambda'}{j}{j'}{I} R_{\alpha,\alpha'}^{Q,\lambda,K}(R) \delta_{I,I'}
\langle J_t,n\|\mathcal{T}_Q(\xi)\|J'_t,n'\rangle,
\end{align}
\end{widetext}
where $\langle J_t,n\|\mathcal{T}_Q(\xi)\|J'_t,n'\rangle$ is the reduced matrix element of $\mathcal{T}_{Qq}$ as defined in Eq.~(\ref{multexpV}) between the states of the target $\lbrace J_t,n\rbrace$, $\lbrace J'_t,n'\rbrace$; $\alpha(\alpha')$ is the final(initial) state of the projectile and includes all quantum numbers necessary for the determination of the corresponding component $\alpha=\lbrace i,l,j,I\rbrace$,
\begin{equation}
 \begin{pmatrix} Q \\ \lambda\end{pmatrix}=\sqrt{\dfrac{(2Q)!}{(2(Q-\lambda))!(2\lambda)!}}
\end{equation}
and $R_{\alpha,\alpha'}^{Q,\lambda,K}(R)$ are defined as follows:
\begin{equation}
\label{eqR}
\begin{split}
R_{\alpha,\alpha'}^{Q,\lambda,K}(R)&=\int \varphi^*_{J_p,\alpha}(r)\varphi_{J_p',\alpha'}(r)\\\times&V^{QK}_{vt}(R,r)  R^\lambda (\gamma_{v}r)^{Q-\lambda} \mathrm{d}r, 
\end{split}
\end{equation}
where $u_{J_p,\alpha}$ are as defined in (\ref{projwf}) and $\alpha$ denotes all the necessary quantum numbers that define the projectile components.
$V^{QK}_{vt}(R,r)$ results from the multipole expansion of $V^{Q}_{vt}$, according to:
\begin{equation}
\label{eqV}
V^{QK}_{vt}(R,r)=\dfrac{1}{2}\int_{-1}^{1} \dfrac{V^{Q}_{vt}(r_{v})}{r_{v}^Q}P_K(u) \mathrm{d}u; \hspace{0.5cm} u=\hat{R}\cdot\hat{r}.
\end{equation}
Similar equations to (\ref{eqR}) and (\ref{eqV}) apply to the core-target interaction, replacing $V_{vt}$ by $V_{ct}$ and $\gamma_v$ by $\gamma_c$. For the following calculations we have chosen the particle-rotor model \citep{bohrmottelson} to describe the collective excitation of the target nucleus. In this model we deform a central potential $V^{(0)} (r)$ through the following transformation:
\begin{equation}
V(\vec{r},\hat{\xi})=V^{(0)}(r-\delta_2Y_{20}(\hat{\xi})),
\end{equation}
and we then perform an expansion in multipoles:
\begin{equation}
V(\vec{r},\hat{\xi})=\sqrt{4\pi}\sum_{Qq}V^Q(r)\mathcal{D}^Q_{q0}(\alpha',\beta',\gamma')Y_{Qq}(\hat{r}),
\end{equation}
where
\begin{equation}
V^Q(r)=\dfrac{\hat{Q}}{2}\int^1_{-1} V^{(0)}(r-\delta_2Y_{20}(\hat{\xi}))P_Q(u) \mathrm{d} u ;\hspace{0.2cm} u=\hat{r}\cdot \hat{\xi},
\end{equation}
$\hat{\xi}$ can be interpreted as the orientation of the axis of the rotor, and $\alpha',\beta', \gamma'$ are the Euler angles that change from the intrinsic frame of the rotor to the laboratory system. 

In this model the reduced matrix element has a simple form:
\begin{equation}
\label{eqT}
\begin{split}
\langle J_t,n\|\mathcal{T}_Q(\hat{\xi})\|J'_t,n'\rangle&=\langle J_t,K\|\mathcal{D}^{Q*}\|J'_t,K\rangle\\&=\hat{J_t}'\langle J'_tKQ0|J_tK\rangle
\end{split}
\end{equation}
It should be noted that, in general, reduced matrix elements for Coulomb and nuclear parts may differ, so it is necessary to compute the contribution of both potentials separately. However, in the particle-rotor model the reduced matrix elements are the same, while the Coulomb potential multipoles are expressed based on the reduced multipole electric operators:

\begin{equation}
V^Q(r)=\dfrac{\sqrt{4\pi}}{\hat{Q}^2}\dfrac{Z_te}{r^{Q+1}}\mathcal{M}(EQ),
\end{equation}

\section{Calculations \label{sec:calc}}

In this section we present some calculations, focusing on the angular differential cross section for the target excitation and its interplay with projectile breakup.

\subsection{\label{sec:58Ni}  $^{58}{\rm Ni}(d,d)^{58}{\rm Ni}^*$} 

As a test of the formalism, we have tried to reproduce the results in \citep{Yahiro01041986} for the reaction $^{58}{\rm Ni}(d,d)^{58}{\rm Ni}^*$ at a deuteron energy of $E_d=80$ $\mathrm{MeV}$. The continuum of the deuteron has been discretized using a pseudo-state method, that is, diagonalising the Hamiltonian in a basis of square-integrable functions, for which we have chosen the transformed harmonic oscillator basis  \citep{moro2009analytical}. After diagonalization, only the states with $p-n$ energies up to 40 MeV and orbital angular momentum $l=0-2$ have been considered, finding higher energies to have a negligible influence on the cross sections.
 We have found the contributions of $l=1,4$ waves to be negligible. The $l=3$ has not been considered since its coupling is expected to be smaller than the $l=1$ coupling, and the latter is already negligible. The target is considered only in its ground ($0^+$) and its first excited ($2^+$) states.

We have obtained the $p-^{58}{\rm Ni}$ and $n-^{58}{\rm Ni}$ potentials from the K$\ddot{{\rm o}}$ning-Delaroche parametrization \citep{Koning2003231} and the $p-n$ interaction is the same from \citep{Yahiro01041986}. The deformation parameter $\beta=0.179$ was chosen for $^{58}{\rm Ni}$ \citep{chartnuclides}, and a rotational model was considered for its deformation. Target excitation due to Coulomb interaction has been neglected, as it is expected to be negligible for these relatively light nuclei. It can be argued that the proper potentials to be used for these calculations are the ``bare'' potentials from which the effect of excitation to the $2^+$ state has been substracted. In order to obtain an approximation to these bare potentials we have calculated new potentials performing a fit to the elastic data using a calculation which explicitly includes excitation to the $2^+$. However, we found the difference between the calculation using these potentials and those from the general parametrization to be small, so we have chosen to keep the potentials from the K$\ddot{\rm o}$ning-Delaroche parametrization for simplicity.

\begin{figure}
{\par\centering \resizebox*{0.45\textwidth}{!}{\includegraphics{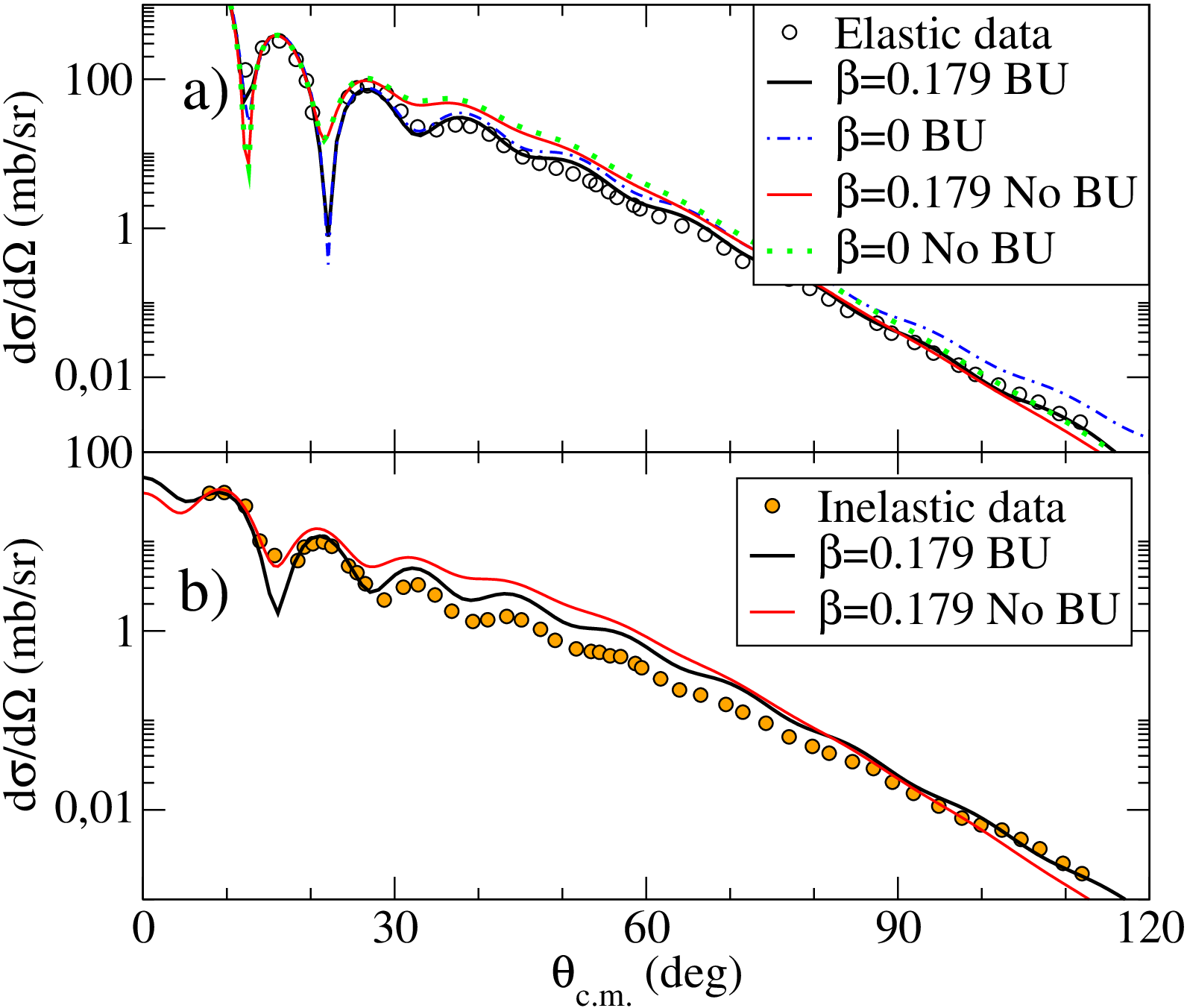}}\par}
 \caption{\label{fig:58Ni} (Colour online) Elastic a) and inelastic b) angular differential cross sections for the $^{58}{\rm Ni}(d,d)^{58}{\rm Ni}$ reaction at a deuteron energy of $E_d=80$ ${\rm MeV}$. The black solid line includes breakup and target excitation consistently. The red solid line only includes target excitation, while the blue dash-dotted line includes only breakup. Finally, the green dotted line excludes both target excitation and breakup. It can be seen that the best agreement with the data is obtained when including both breakup and target excitation effects.} 
\end{figure}

The results are shown in Fig.~\ref{fig:58Ni} for the elastic (top) and inelastic (bottom ) along with the experimental data \citep{stephenson1980proceedings}. It is seen that the calculations are in good agreement with the data. Moreover, the best agreement is obtained when including both breakup and target excitation effects.

\subsection{\label{sec:24Mg}  $^{24}{\rm Mg}(d,d)^{24}{\rm Mg}^*$} 

\begin{figure}
{\par\centering \resizebox*{0.45\textwidth}{!}{\includegraphics{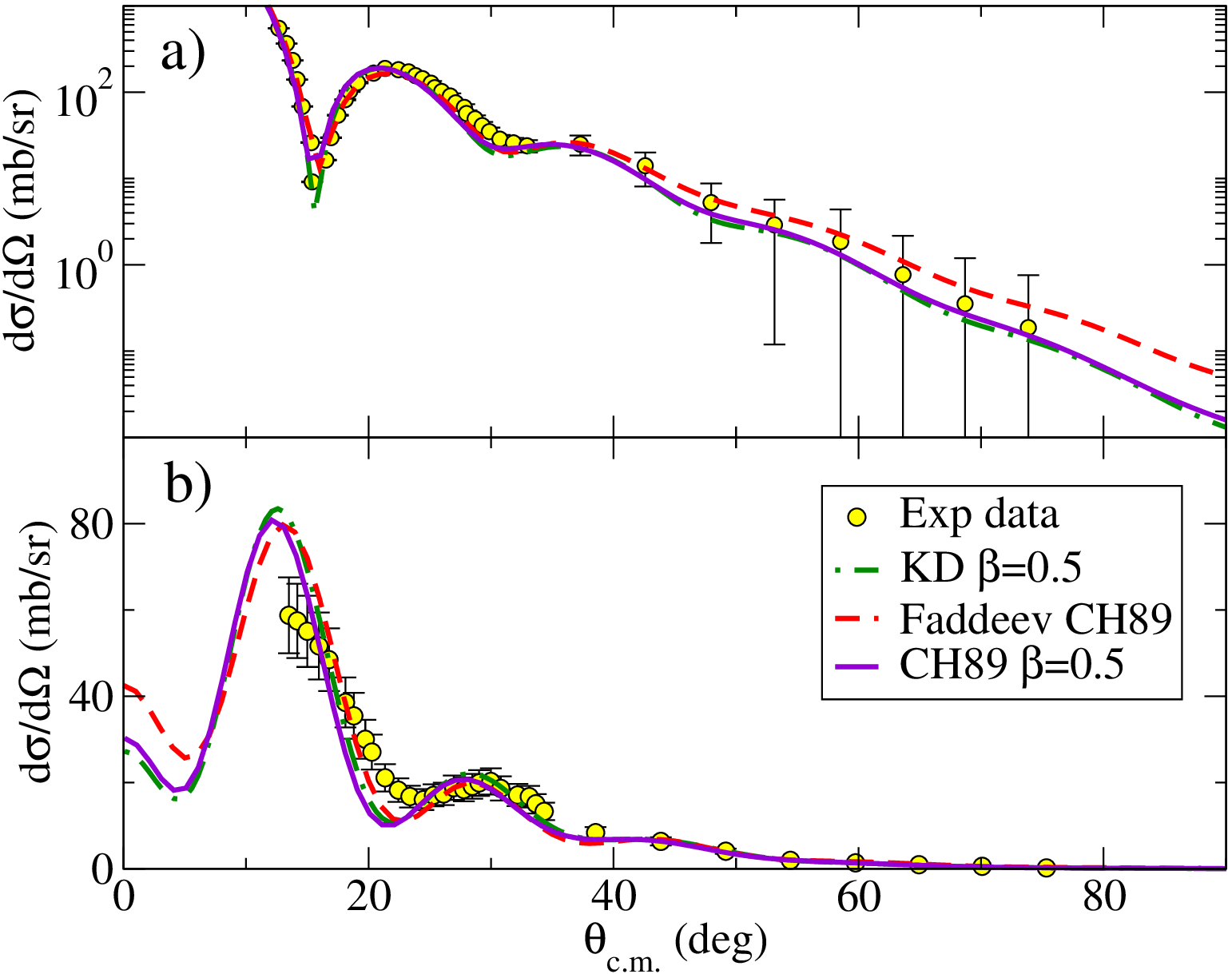}}\par}
 \caption{\label{fig:24Mg} (Colour online) Elastic (upper) and inelastic (lower) differential angular cross sections for $^{24}{\rm Mg}(d,d)^{24}{\rm Mg}^*$ at a deuteron energy of $70$ ${\rm MeV}$. CDCC calculations are presented using K$\ddot{\mathrm{o}}$ning-Delaroche and CH89 parametrizations. Faddeev calculations are taken from \citep{Deltuva2016173}.}
\end{figure}

In this section we study the reaction $^{24}{\rm Mg}(d,d)^{24}{\rm Mg}^*$ leading to the ground and first excited states of $^{24}{\rm Mg}$. This reaction was measured at different energies in the range of tens of MeV \citep{Kiss19761}, from which we will focus on the reaction at a deuteron energy of $70$ ${\rm MeV}$. For the deformation parameter of $^{24}{\rm Mg}$ we have chosen a value $\beta=0.5$, in accordance to \citep{Deltuva2016173}. We have performed calculations using K$\ddot{\mathrm{o}}$ning-Delaroche \citep{Koning2003231} and CH89 \citep{CH89} parametrizations for $p,n-^{24}{\rm Mg}$ and the potential from \citep{Yahiro01041986} for the $p-n$ interaction. The results of our calculations, as well as those obtained from Faddeev calculations \citep{Deltuva2016173} employing CH89 parametrization and $p-n$ CD Bonn potential \citep{Bonn} are presented in Fig.~\ref{fig:24Mg}. Although the calculations differ slightly, all seem to agree equally well with experimental data. It seems as though a smaller deformation parameter would give a better fit to the experimental data, as indicated in \citep{Deltuva2016173}. However, in this same paper different deformation parameters were found to give best fits to data at different deuteron energies. Therefore, in this paper we will not try to extract a value for the deformation parameter of $^{24}{\rm Mg}$ beyond confirming the conclusions in \citep{Deltuva2016173}.

\begin{figure}
{\par\centering \resizebox*{0.45\textwidth}{!}{\includegraphics{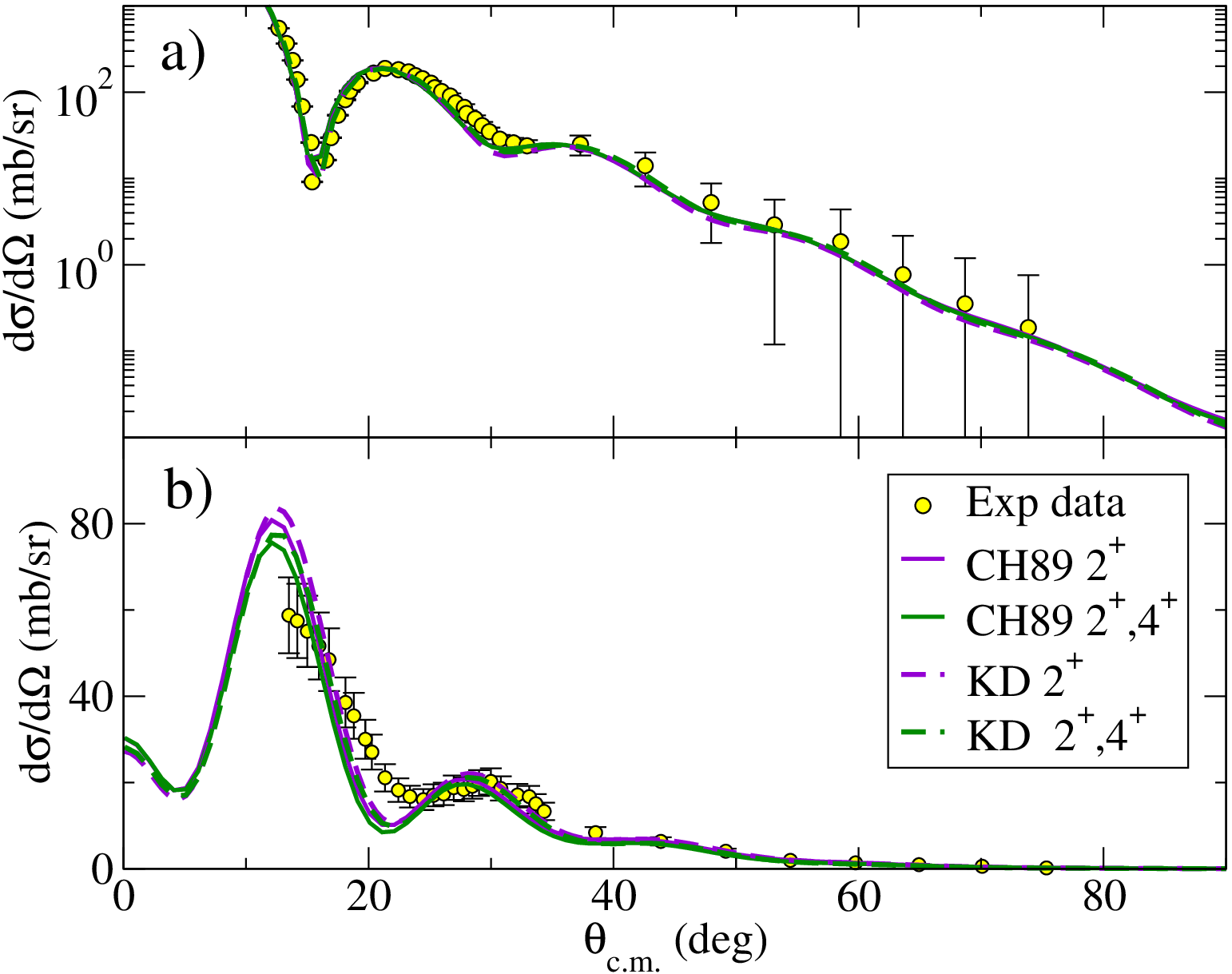}}\par}
 \caption{\label{fig:24Mg4+} (Colour online) Elastic (upper) and inelastic (lower) differential angular cross sections for $^{24}{\rm Mg}(d,d)^{24}{\rm Mg}^*$ at a deuteron energy of $70$ ${\rm MeV}$ computed including (violet lines) and excluding (green lines) the $4^+$ state of $^{24}{\rm Mg}$, using CH89 (solid lines) and K$\ddot{\mathrm{o}}$ning-Delaroche (dashed lines) parametrizations.}
\end{figure}

In \citep{Deltuva2016173} the inclusion of higher excited states of $^{24}{\rm Mg}$ was suggested but was not calculated due to the computational cost of the Faddeev calculation. Since the method used here is less demanding computationally than that used in \citep{Deltuva2016173}, we have performed calculations including the $2^+$ and $4^+$ states of $^{24}{\rm Mg}$, coupled through quadrupolar and hexadecapolar deformation $\beta_4=-0.017$ \citep{PhysRevC.23.1355}. The results are shown in Fig.~\ref{fig:24Mg4+}. The effect of the $4^+$ state is found to be rather small, being unnoticeable in the elastic cross section and leading to a small reduction of the cross section at the peak of the inelastic cross section.

\subsection{\label{sec:144Sm}  $^{144}{\rm Sm}(^6{\rm Li},^6{\rm Li})^{144}{\rm Sm}^*$} 

In this section we study the inelastic scattering of $^6{\rm Li}$ from $^{144}{\rm Sm}$ which has been recently measured at energies around the Coulomb barrier \citep{Woodard201217}. This reaction has been previously studied \citep{Woodard201217} by decoupling target excitation and breakup of the projectile. First, CDCC was used to calculate breakup and a folding potential was obtained from this calculation. This potential was then deformed to account for target excitation.
Due to this decoupling, target excitation and breakup of the projectile are not treated consistently. Since \nuc{6}{Li} is relatively weakly bound, its breakup is an important process that can couple to the excitation of the collective degrees of freedom of \nuc{144}{Sm}, so a consistent treatment of both effects is of importance to describe this reaction. The extension of CDCC described in Sec. \ref{sec:theory} allows to treat both processes on equal footing. Therefore, we find its application to this reaction to be suitable and relevant.

We focus on the elastic and inelastic ($^{144}{\rm Sm}(^6{\rm Li},^6{\rm Li})^{144}{\rm Sm}^*$) angular differential cross section as our observables of interest. Experimental data exist for four different incoming \nuc{6}{Li} energies: $23$, $28$, $30$ and $35$ ${\rm MeV}$. This allows for a certain systematic study of the results obtained.

For the following calculations we have considered \nuc{6}{Li} to be a binary system composed of an inert deuteron and an inert $\alpha$ particle, so that their excitations and possible breakup are neglected. For \nuc{144}{Sm} only the first $2^+$ and $3^-$ states have been considered, with excitation energies of $1.66$ ${\rm MeV}$ and $1.81$ ${\rm MeV}$ respectively. The deformation parameters $\beta_2=0.087$ and $\beta_3=0.130$ \citep{Woodard201217} have been used to deform the $d/\alpha-^{144}{\rm Sm}$ potentials. For this reaction, the effect of Coulomb can no longer be neglected, as it was the case in the previous sections. Therefore, it is necessary to include the possibility of target excitation through Coulomb interaction, for which the reduced Coulomb matrix elements must be obtained. We have computed the matrix elements from the $B(E\lambda)$ probabilities using \citep{thompsonnunes} Eq. (4.4.3):

\begin{equation}
B(E\lambda, I_i\rightarrow I_f)=\dfrac{1}{2I_i+1}|\left\langle I_f\| E\lambda \| Ii \right\rangle|^2,
\end{equation}
using the $B(E\lambda, I_i\rightarrow I_f)$ values from \citep{RAMAN20011,KIBEDI200235}.

The continuum of \nuc{6}{Li} has been discretized using a transformed harmonic oscillator basis \citep{PhysRevC.71.064601} extending up to 10 MeV and considering only the $s$ and $d$ waves. The potential used for the $d-\alpha$ interaction has been chosen of a Gaussian shape \citep{PhysRevC.68.064607} and its depth has been adjusted for the $s$ wave to obtain the ground state at $E=-1.47$ ${\rm MeV}$ and for the $d$ wave to obtain a resonance at $E=0.7$ ${\rm MeV}$, to reproduce the experimental $3^+$ resonance. The width parameter used is $a=2.236$ fm while the depths of the potential are $V_0=74.88$ MeV and $V_2=85.30$ MeV, for $l=0,2$ respectively. The spin of the deuteron has been neglected throughout the following calculations. 
Due to the experimental impossibility of distinguishing between the $2^+$ and $3^-$ states of \nuc{144}{Sm} the inelastic cross sections presented are the sum of the contributions for both states.

\begin{figure}
{\par\centering \resizebox*{0.5\textwidth}{!}{\includegraphics{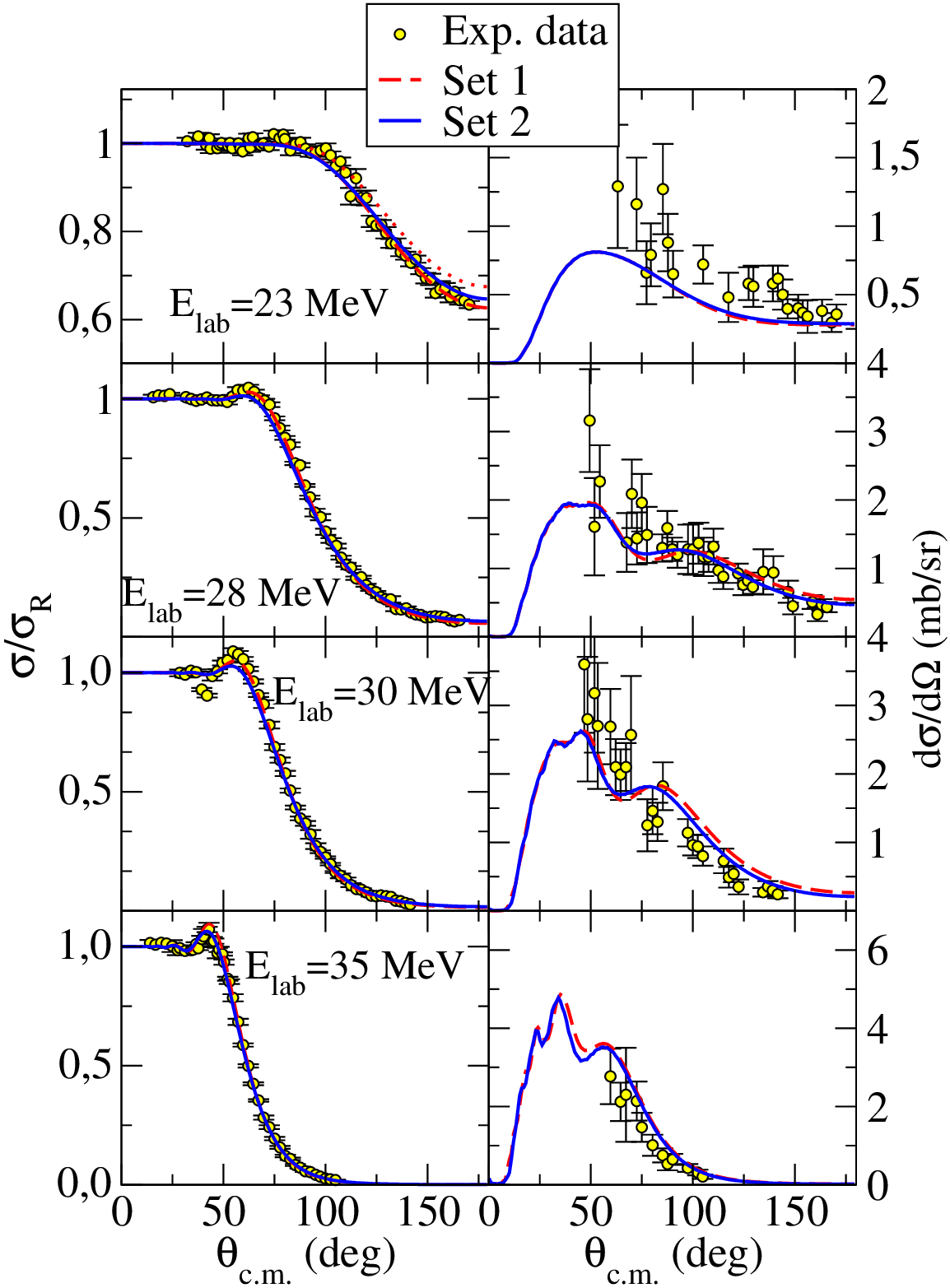}}\par}
 \caption{\label{fig:6Lipot} (Colour online) Elastic (left) and inelastic (right) differential angular cross sections for $^{144}{\rm Sm}(^6{\rm Li},^6{\rm Li})^{144}{\rm Sm}^*$ for incoming energies of  $23,$ $28,$ $30$ and $35$ ${\rm MeV}$. The red dashed and solid blue lines correspond to different $\alpha-^{144}{\rm Sm}$ and $d-^{144}{\rm Sm}$ potentials (see text). For $23$ ${\rm MeV}$, the elastic cross section using a renormalization factor of $N_{id}=0.2$ is shown with a red dotted line (see text).}
\end{figure}

For the $d-^{144}{\rm Sm}$ and $\alpha-^{144}{\rm Sm}$ potentials different parametrizations have been used. In Fig.~\ref{fig:6Lipot} we present calculations employing two different sets of potentials.

 The red dashed line (Set 1) potential has been constructed using the Sao Paulo parametrization \citep{PhysRevLett.79.5218}, both for the $\alpha-^{144}{\rm Sm}$ and $d-^{144}{\rm Sm}$ interactions. Since the energies considered are not too different, the dependence of the potential on the energy has been neglected. The imaginary part of the potential was rescaled by a factor $N_{i\alpha}$ for the $\alpha-^{144}{\rm Sm}$ potential and by a factor $N_{id}$ for the $d-^{144}{\rm Sm}$ potential, in order to reproduce the elastic scattering data. A fairly negligible dependence on $N_{i\alpha}$ was found, therefore a value of $N_{i\alpha}=1$ was chosen. For the $d-^{144}{\rm Sm}$ potential, we could not find a renormalization value that would give a reasonable agreement with the experimental data for all incident energies: We found that a renormalization value of $N_{id}=0.2$ agrees well with the data for incident energies of $28,$ $30 $ and $35$ ${\rm MeV}$. However for $23$ ${\rm MeV}$, a factor of $N_{id}=0.5$ was needed to give a good agreement. In Fig.~\ref{fig:6Lipot} the dotted red line represents calculations for $E_{^6Li}=23$ MeV using a renormalization factor of $N_{id}=0.2$.
 
 The solid blue line (Set 2) corresponds to a calculation in which the $\alpha-^{144}{\rm Sm}$ potential is based once again on the Sao Paulo parametrization. The imaginary potential has been rescaled using a factor of $N_{i\alpha}=0.78$ this time. Meanwhile, for the $d-^{144}{\rm Sm}$ potential, we have chosen Perey-Perey global potential \citep{PhysRev.132.755}, whose imaginary part has been rescaled using a factor of $N_{id}=0.7$. This reduction of the imaginary potential for the deuteron inside \nuc{6}{Li} has been discussed previously \citep{Hirabayashi199111, Santra2009139, PhysRevC.75.054605, PhysRevC.86.031601}. We find this selection of potentials to give a remarkably good fit to the data at all considered incident energies.

In general we find the inelastic cross section to be quite insensitive to the selection of potentials, while Set 1 seems to give a better agreement with the elastic data around the Coulomb-nuclear interference peak.

\begin{figure}
{\par\centering \resizebox*{0.5\textwidth}{!}{\includegraphics{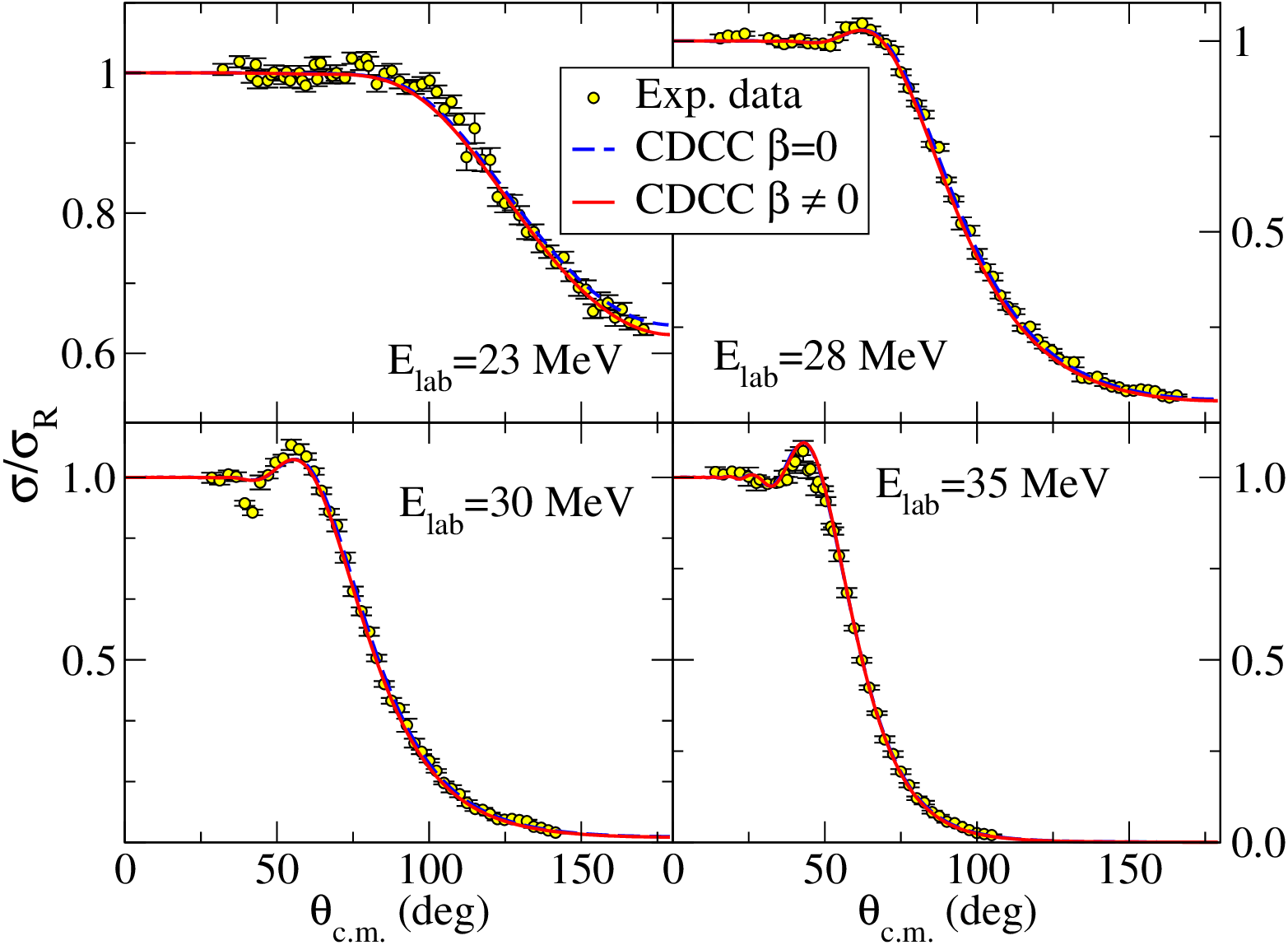}}\par}
 \caption{\label{fig:6LiNodef} (Colour online) Elastic angular differential cross section for different incident energies for the reaction $^{144}{\rm Sm}(^6{\rm Li},^6{\rm Li})^{144}{\rm Sm}$. The red solid line corresponds to calculations using Set 1 (see text) of potentials including deformation of the target, while the blue dashed line corresponds to calculations using the same potentials but without target deformation.}
\end{figure}

We find that the inclusion of target excitation in the calculations gives a small contribution to the elastic cross section. This can be seen in Fig.~\ref{fig:6LiNodef}, where the solid red line corresponds to calculations using potentials from Set 1, described above, including deformation of the target. Meanwhile, the blue dashed line represents calculations using the same potentials but setting the deformation parameter to 0. As can be seen, both calculations give very similar results.

\begin{figure}
{\par\centering \resizebox*{0.35\textheight}{!}{\includegraphics{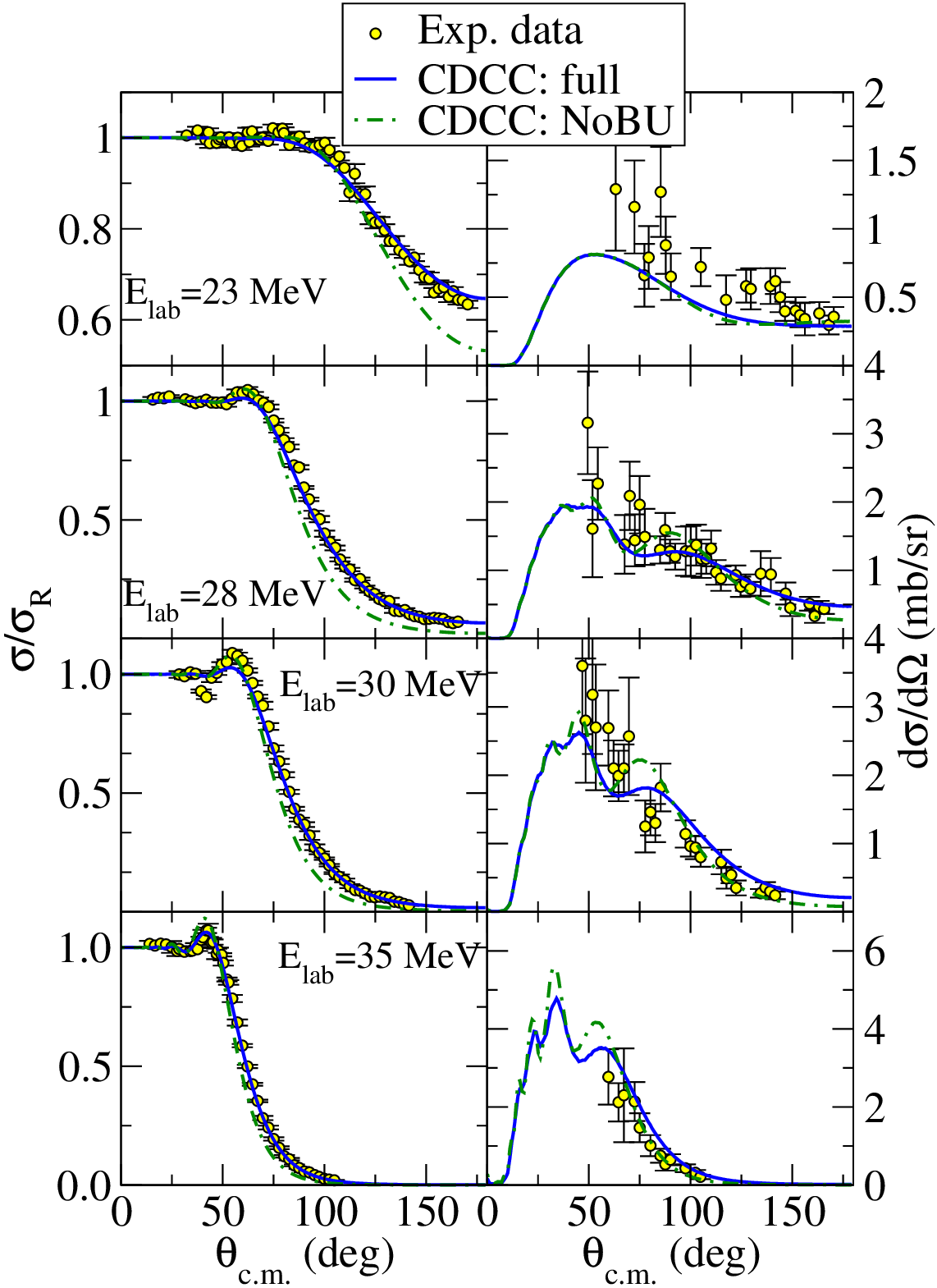}}\par}
 \caption{\label{fig:6LiNoBU} (Colour online) Elastic (left) and inelastic (right) differential cross sections for $^{144}{\rm Sm}(^6{\rm Li},^6{\rm Li})^{144}{\rm Sm}^*$ at different incident energies. Coupled-channel calculations are presented using potentials from Set 2 (see text) including (solid blue line) and excluding (dash-dotted green line) breakup states from the calculation.}
\end{figure}

In order to study the effect of breakup in both elastic and inelastic cross sections we have performed a calculation excluding all breakup states from the coupled-channel calculation, for which we have chosen the potentials from Set 2. The results are shown in Fig.~\ref{fig:6LiNoBU}. As can be seen in the figure, without breakup, the elastic differential cross section is underestimated at all the energies at larger angles, probably due to the exclusion of the resonance at $E_{d\alpha}=0.7$ MeV. We can see also that the Fresnel peak is displaced, specially for the incident energy of $35$ MeV. As for the inelastic cross section, it is found that the effect of breakup is only moderate, smoothing the oscillation of the cross section, but not modifying its magnitude much. It can be seen that the effect of the breakup states becomes more important with increasing incident energy but even for the highest energy the agreement with the data of both calculations is similar.

\begin{figure}
{\par\centering \resizebox*{0.35\textheight}{!}{\includegraphics{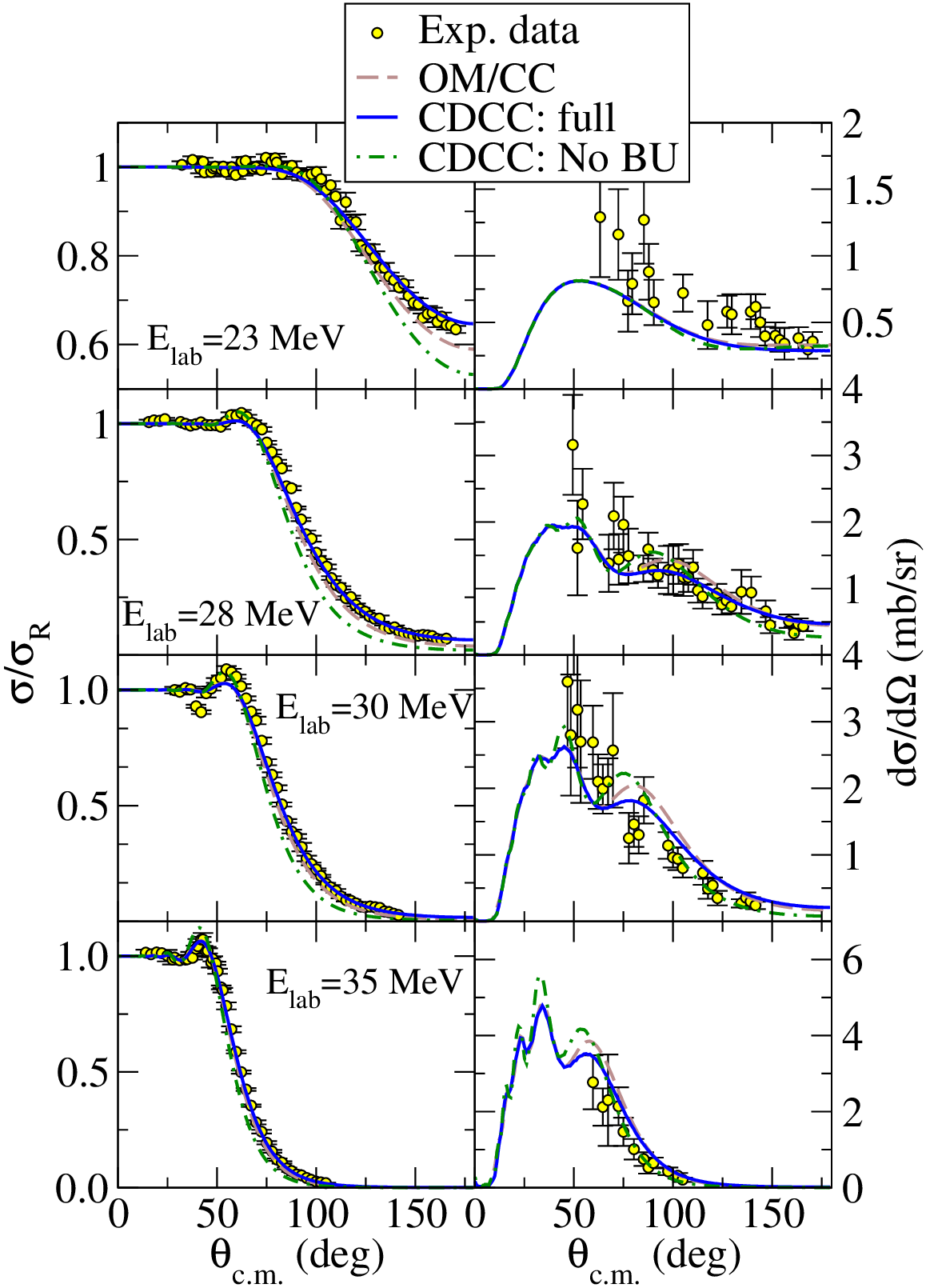}}\par}
 \caption{\label{fig:6LiCook} (Colour online) Elastic (left) and inelastic (right) differential cross sections for $^{144}{\rm Sm}(^6{\rm Li},^6{\rm Li})^{144}{\rm Sm}^*$ at different incident energies. Coupled-channel calculations are presented using potentials from Set 2 (see text) including (solid blue line) and excluding (dash-dotted green line) breakup states from the calculation. The dashed brown line corresponds to optical model calculations using Cook potential \citep{Cook1982153}.}
\end{figure}

To end with the study of this reaction, we have performed a coupled-channel calculation in which the incident channel is described by means of an optical potential and so breakup effects are only accounted for effectively. We have used Cook potential \citep{Cook1982153}, which we have deformed employing the deformation parameters indicated above, assuming a rotor model for \nuc{144}{Sm}. The results are shown in Fig.~\ref{fig:6LiCook}, together with the curves of Fig.~\ref{fig:6LiNoBU}, in order to compare it with the full CDCC calculation and the calculation without breakup. It is rather remarkable the good agreement that is obtained for the inelastic cross section between the full CDCC calculation and the optical model one: the curves overlap except for a small increase at intermediate angles for the optical model result. Let us remark that both calculations use potentials from completely different systematics, the only point in common being the deformation parameters and the model used for the deformation.
In the elastic cross section we can see that the optical model calculation seems to underestimate the data at the large-angle tail probably due once again to an inadequate treatment of the \nuc{6}{Li} resonance.

\subsection{\label{sec:9Be}  $^{9}{\rm Be}(^6{\rm Li},^6{\rm Li})^{9}{\rm Be}^*$} 

In this last section, we have performed calculations for the reaction $^{9}{\rm Be}(^6{\rm Li},^6{\rm Li})^{9}{\rm Be}$ at an incident energy of 20 MeV. We present calculations for the cross sections of elastic scattering; excitation to the excited state of \nuc{9}{Be}, with excitation energy $E_x=2.43$ MeV and angular momentum $J^\pi=5/2^-$, and excitation to the resonant state of \nuc{6}{Li} with $E_x=2.19$ MeV and $J^\pi=3^+$, for which experimental data exist \citep{MUSKAT199542}.

For these calculations we have considered a rotor model for \nuc{9}{Be}, for which a deformation length of $\delta=2.5$ fm has been chosen, following \citep{MUSKAT199542}. \nuc{6}{Li} is treated as a binary system composed of a deuteron and an $\alpha$ particle as in the previous section. However, in order to obtain the correct magnitude of the cross section corresponding to the excitation of \nuc{6}{Li}, it has been necessary to include the spin of the deuteron, in contrast to the previous section's calculations.

The potentials used in the calculations are indicated as follows: for the $\alpha-d$ interaction the potential of \citep{PhysRevC.68.044607}, which includes the deuteron spin and gives a proper description of both the ground state and the $3^+$ resonance of \nuc{6}{Li}, has been used.

For the $\alpha-^9$Be interaction, the potential from \citep{Taylor1965318} has been chosen. Due to the difficulty of including the spin-orbit of the fragment-target interaction in the CDCC calculations, the $\alpha-^9$Be spin-orbit term has been applied to the whole \nuc{6}{Li}-\nuc{9}{Be} system, following the prescription of \citep{PhysRevC.83.064603}.

\begin{figure}
{\par\centering \resizebox*{0.35\textheight}{!}{\includegraphics{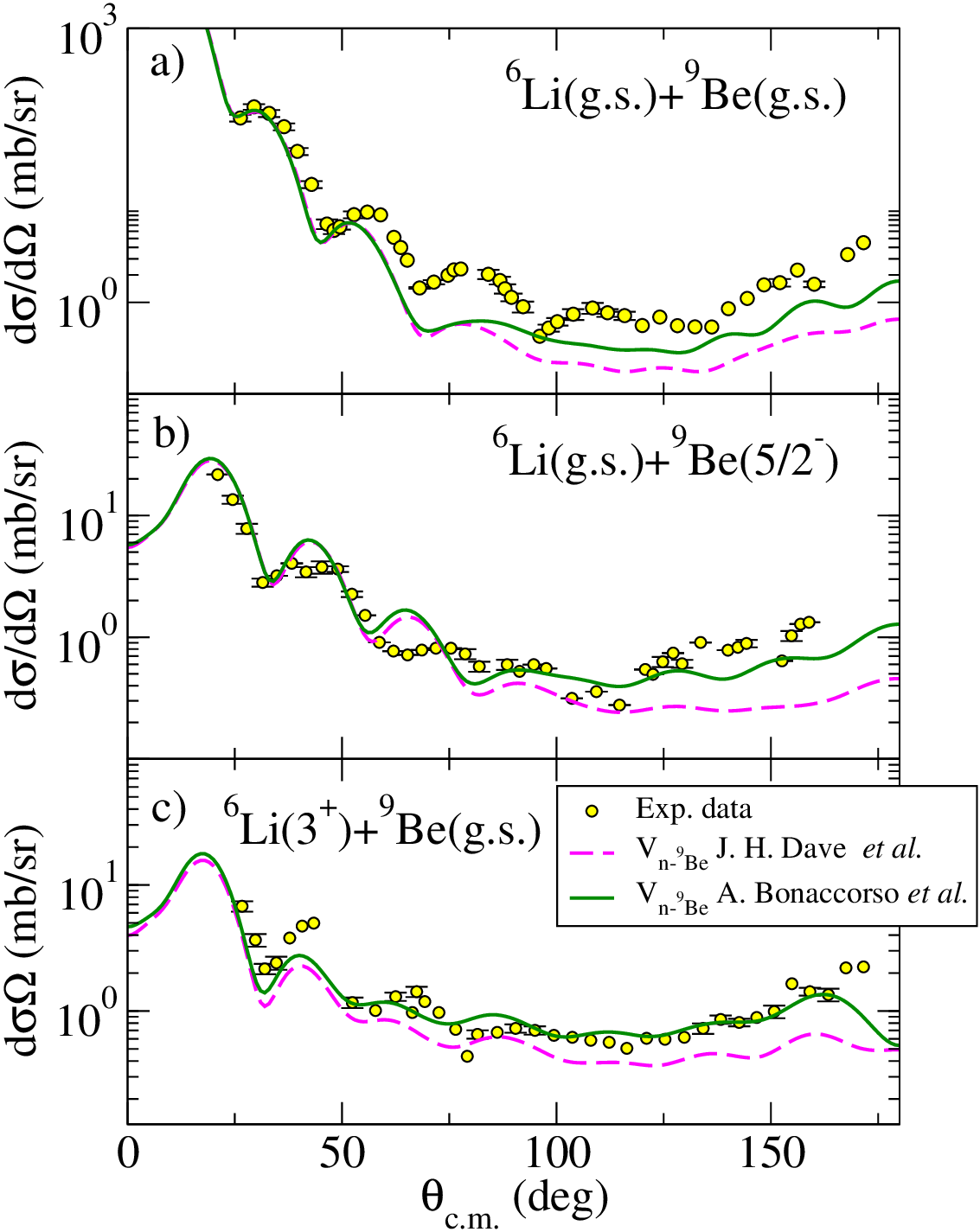}}\par}
 \caption{\label{fig:9Be1} (Colour online) Cross sections for elastic scattering a) and  excitation of $^9$Be ($E_x=2.43 \mathrm{ MeV}, J^\pi=5/2^-$) b) and of $^6$Li (resonant state at $E_x=2.19 \mathrm{ MeV}$, $J^\pi=3^+$) c). Two sets of potentials are presented (see text).}
\end{figure}

Finally, the $d-^9$Be potential has been obtained by folding the ground state of the deuteron with $p-^9$Be and $n-^9$Be interactions. For the $p-^9$Be interaction the potential of \citep{PhysRevC.28.2212} has been used. Although this potential was derived for neutrons, we have found a small dependence of the observables on it. Therefore, this prescription has been chosen for convenience. For the $n-^9$Be potential, two prescriptions have been used. In Fig. \ref{fig:9Be1}, the magenta dashed line corresponds to $V_{n^9\mathrm{Be}}$ from \citep{PhysRevC.28.2212}, while the green solid line corresponds to the parametrization from \citep{PhysRevC.89.024619}, which gives a more precise description of the $n-^{9}$Be system at low energies. It is seen that both potentials give similar cross sections up to $50-60^\circ$, but differ afterwards finding that the potential from \citep{PhysRevC.89.024619} gives a better agreement with the experimental data.

We find a reasonably good agreement between the calculations and the experimental data for both the excitation of \nuc{9}{Be} and \nuc{6}{Li}. However, we find an underestimation of the elastic data for larger angles. This could be due to the contribution of other channels, such as compound nucleus and cluster transfer, that were suggested in \citep{MUSKAT199542} but are not considered here. It must be noted that all observables have been obtained using a consistent calculation including breakup of \nuc{6}{Li} and excitation of \nuc{9}{Be}.

Since the main purpose of this paper is the portrayal of the formalism for target excitation within CDCC and because of the general difficulty of treating \nuc{6}{Li} within the two-cluster formalism for reactions with light nuclei \citep{Hirabayashi199111}, we have decided not to pursue a better agreement between our calculations for elastic scattering and experimental data, deeming it beyond the scope of the present article.

\begin{figure}
\includegraphics[width=0.4\textwidth]{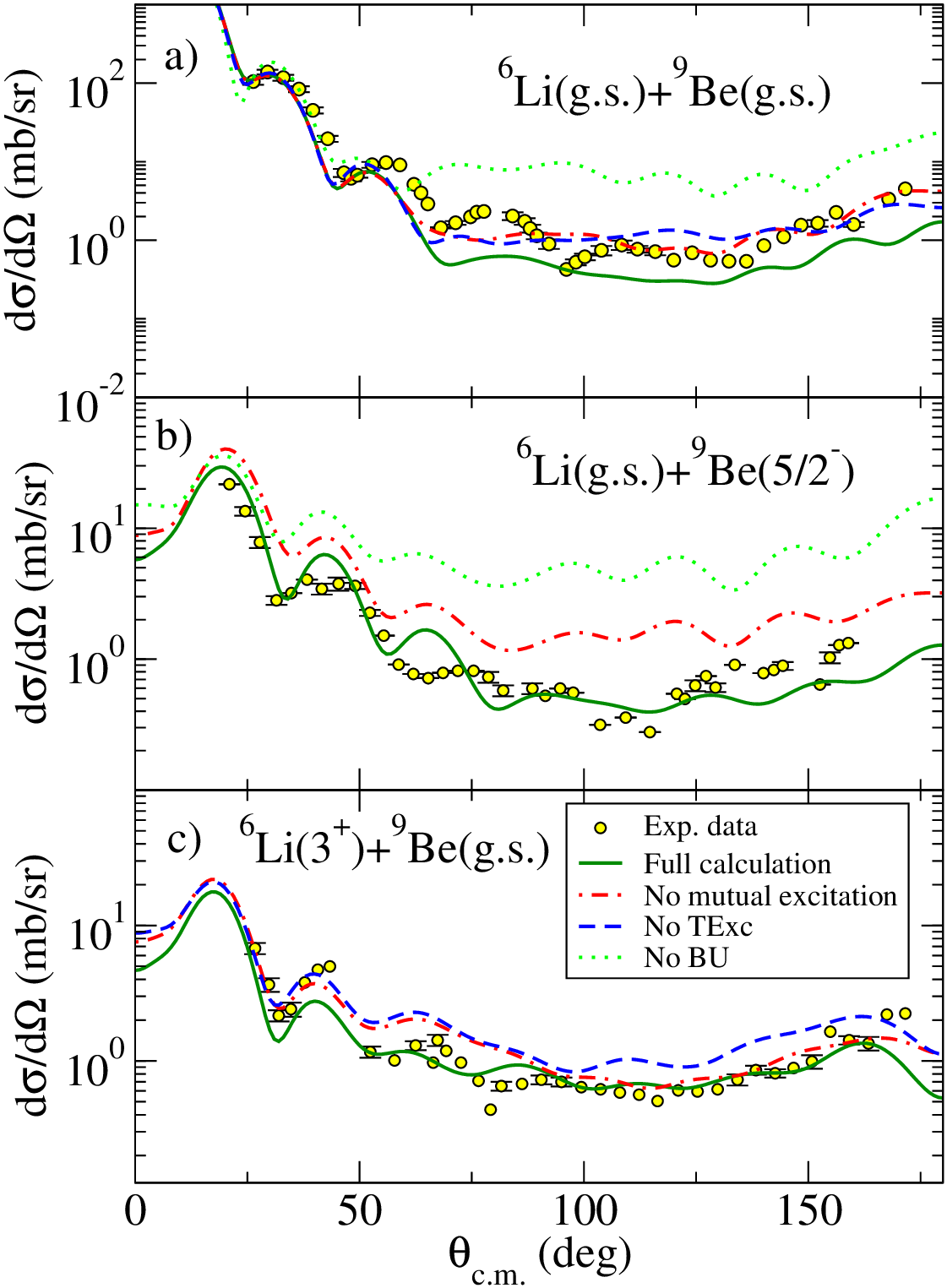}
 \caption{\label{fig:9Be2} (Colour online) Cross sections for elastic scattering a) and inelastic scattering of $^9$Be ($E_x=2.43 \mathrm{ MeV}, J^\pi=5/2^-$) b) and of $^6$Li (resonant state at $E_x=2.19 \mathrm{ MeV}$, $J^\pi=3^+$) c). Calculations have been performed using the potential from \citep{PhysRevC.89.024619} for $V_{n-^9\mathrm{Be}}$. The rest of the potentials are those indicated in the text. The dark-green solid line corresponds to the full calculation, while the red dot-dashed line excludes states with both \nuc{9}{Be} and \nuc{6}{Li} in their excited states. The blue dashed line excludes all states with \nuc{9}{Be} in its excited state while the green dotted line excludes all breakup states of \nuc{6}{Li}.}
\end{figure}

In order to study the effect of breakup, target excitation and their interplay we have performed calculations excluding different channels. The results are shown in Fig. \ref{fig:9Be2}. In it the dark-green solid line corresponds to the full calculation, while for the red dot-dashed line the states in which \nuc{9}{Be} is in its excited state and \nuc{6}{Li} is in a continuum state at the same time have been excluded. For the dashed blue line, the excitation of \nuc{9}{Be} has been omitted while for the dotted green line no breakup states of \nuc{6}{Li} have been included.

It is remarkable that the effect of the states where both \nuc{9}{Be} and \nuc{6}{Li} are excited (which we may call mutual excitation) is quite important for all cross sections presented, as can be seen on the red dot-dashed line, which gives an increase in all cross sections, particularly in the cross section corresponding to the excitation of \nuc{9}{Be}. 
While the effect of the mutual-excitation states would be expected to be small, it is however found to be quite relevant to the cross section, smaller than the effect of the direct breakup of \nuc{6}{Li} (as can be seen comparing the green dotted and red dot-dashed lines) but quite more relevant than the excitation of \nuc{9}{Be} without simultaneous breakup of \nuc{6}{Li}, as is highlighted by the similarity between the red dot-dashed and blue dashed lines.

\section{\label{sec:summary} Summary and conclusions}

The extension of the CDCC method to include excitation of collective degrees of freedom of the target has been developed without any approximation apart from those inherent to the CDCC method. The formalism has been implemented in a computer code and applied to a number of reactions. 

For the reaction $d$+\nuc{24}{Mg} our results reproduce very well those obtained with the more sophisticated Faddeev/AGS method. For \nuc{6}{Li}+\nuc{144}{Sm} the calculations reproduce very nicely the inelastic cross sections at four different energies, although the reproduction of the elastic scattering data required a readjustment of the deuteron-target potential, in accordance to previous findings. Finally, for the reaction \nuc{6}{Li}+\nuc{9}{Be} the effects of the interplay of target excitation and projectile breakup seem to be more important than in the previous reactions, probably due to the lighter nuclei and the inferior beam energy, which leads to a longer interaction time. 

The method developed here has potential applications in several problems of timely interest. For example, it could be used to study the simultaneous influence of projectile breakup and collective excitations of the target on fusion cross sections in reactions induced by weakly-bound projectiles on deformed targets.
\begin{acknowledgments}
We are thankful to A. Deltuva for providing the Faddeev calculations for Fig. \ref{fig:24Mg} and to J. Lubian for calculating the Sao Paulo potentials used in Sec. \ref{sec:144Sm}. We would also like to thank A.~Di Pietro for fruitful discussions.
This work has been partially supported by the Spanish Consolider-Ingenio 2010 Programme CPAN (CSD2007-00042), by Junta de  Andaluc\'ia (FQM160, P07-FQM-02894) and by Spanish Ministerio de Econom\'ia y Competitividad (FIS2014-53448-C2-1-P). M.G.-R.\ acknowledges a research grant by the Spanish
 Ministerio de Educaci\'on, Cultura y Deporte, FPU Research Grant FPU13/014109.
\end{acknowledgments}
\appendix

\section{$Q=0$ reduction}

In this appendix, we calculate the expressions corresponding to $Q=0$ from the formulae presented in Sec.~\ref{sec:theory} and prove that they reduce to <the standard CDCC expressions without target excitation as can be found for example in \citep{fresco}. It must be noted that, when $Q=0$ there is no target excitation but the projectile can still break in this formalism.

We start by noting that Eq.~(\ref{eqV}) and (\ref{eqR}) are reduced to:
\begin{align}
V^{0K}_{vt}(R,r)=\dfrac{1}{2}\int_{-1}^{1} V_{vt}(r_{v})P_K(u) \mathrm{d}u, \\
R_{\alpha,\alpha'}^{0,0,K}(R)=\int \varphi^*_{J_p,\alpha}(r)V^{0K}_{vt}(R,r) \varphi_{J_p',\alpha'}(r) \mathrm{d}r,
\end{align}
using the notation described in Sec. \ref{sec:theory} and the fact that $\lambda \leq Q$.

Next we perform the reduction of (\ref{eqF}) using the properties of $3j$ and $6j$ symbols:
\begin{align}
\tj{0}{A}{B}{0}{0}{0}=(-1)^A\dfrac{\delta_{AB}}{\hat{A}} \\
\sj{A}{B}{C}{D}{E}{0}=(-1)^{A+B+C}\dfrac{\delta_{AE}\delta_{BD}}{\hat{A}\hat{B}},
\end{align}
and the reduction of (\ref{eqT})
\begin{equation}
\langle J_t,n\|\mathcal{T}_0(\hat{\xi})\|J'_t,n'\rangle=\hat{J_t}'\langle J'_tK00|J_tK\rangle=\hat{J_t}'\delta_{J_t'J_t},
\end{equation}
which also indicates that for every model used to characterise the target we must require
$\langle J_t,n\|\mathcal{T}_0(\hat{\xi})\|J'_t,n'\rangle=\hat{J_t}'\delta_{J_t'J_t}$. The reduction yields:
\begin{align}
F^{K}_{\beta\beta'} (R)&=(-1)^{J_p'+K}\sqrt{4\pi}\delta_{\Lambda,\Lambda',K}\hat{J_p}\hat{J_p'}\hat{l}\hat{l}'\hat{j}\hat{j'}\hat{K}^2\hat{J_t}'
\nonumber\\\times&	\sum_{\alpha,\alpha'}(-1)^{j'+j+l+l'+s+I}\tj{l'}{l}{K}{0}{0}{0}
\nonumber\\\times&\sj{j'}{j}{K}{l}{l'}{s}
\sj{J_p'}{J_p}{K}{j}{j'}{I} R_{\alpha,\alpha'}^{0,0,K}(R) \delta_{I,I'}.
\end{align}
Now for the reduction of (\ref{eqV}) we need the reduction of 9j symbols:
\begin{equation}
\nj{A}{B}{C}{D}{E}{0}{F}{G}{H}=(-1)^{G+A+C+D}\dfrac{\delta_{DE}\delta_{CH}}{\hat{D}\hat{C}}\sj{A}{B}{C}{G}{F}{D},
\end{equation}
leading to:
\begin{align}
U^{J_T}_{\beta,\beta'} (R)&=\sum_{K,K,0}\delta_{JJ'}\delta_{J_tJ_t'}\delta_{\Lambda,\Lambda',K}(-1)^{J_p+J+L'+L}\hat{L}\hat{L}'
\nonumber\\\times&\dfrac{1}{\sqrt{4\pi}}\sj{L}{J_p}{J}{J_p'}{L'}{K} \tj{L}{L'}{K}{0}{0}{0}
F^{K}_{\beta\beta'} (R).
\end{align}
Both of these expressions can be shown to be equivalent to those found in \citep{fresco}, although the comparison is not straightforward and requires some angular momentum algebra.
\bibliography{transfer}

\begin{thebibliography}{50}
\expandafter\ifx\csname natexlab\endcsname\relax\def\natexlab#1{#1}\fi
\expandafter\ifx\csname bibnamefont\endcsname\relax
  \def\bibnamefont#1{#1}\fi
\expandafter\ifx\csname bibfnamefont\endcsname\relax
  \def\bibfnamefont#1{#1}\fi
\expandafter\ifx\csname citenamefont\endcsname\relax
  \def\citenamefont#1{#1}\fi
\expandafter\ifx\csname url\endcsname\relax
  \def\url#1{\texttt{#1}}\fi
\expandafter\ifx\csname urlprefix\endcsname\relax\def\urlprefix{URL }\fi
\providecommand{\bibinfo}[2]{#2}
\providecommand{\eprint}[2][]{\url{#2}}

\bibitem[{\citenamefont{Rawitscher}(1974)}]{PhysRevC.9.2210CDCC}
\bibinfo{author}{\bibfnamefont{G.~H.} \bibnamefont{Rawitscher}},
  \bibinfo{journal}{Phys. Rev. C} \textbf{\bibinfo{volume}{9}},
  \bibinfo{pages}{2210} (\bibinfo{year}{1974}).

\bibitem[{\citenamefont{Austern et~al.}(1987)\citenamefont{Austern, Iseri,
  Kamimura, Kawai, Rawitscher, and Yahiro}}]{Austern1987125CDCC}
\bibinfo{author}{\bibfnamefont{N.}~\bibnamefont{Austern}},
  \bibinfo{author}{\bibfnamefont{Y.}~\bibnamefont{Iseri}},
  \bibinfo{author}{\bibfnamefont{M.}~\bibnamefont{Kamimura}},
  \bibinfo{author}{\bibfnamefont{M.}~\bibnamefont{Kawai}},
  \bibinfo{author}{\bibfnamefont{G.}~\bibnamefont{Rawitscher}},
  \bibnamefont{and} \bibinfo{author}{\bibfnamefont{M.}~\bibnamefont{Yahiro}},
  \bibinfo{journal}{Physics Reports} \textbf{\bibinfo{volume}{154}},
  \bibinfo{pages}{125 } (\bibinfo{year}{1987}), ISSN \bibinfo{issn}{0370-1573}.

\bibitem[{\citenamefont{Banerjee and Shyam}(2000)}]{PhysRevC.61.047301}
\bibinfo{author}{\bibfnamefont{P.}~\bibnamefont{Banerjee}} \bibnamefont{and}
  \bibinfo{author}{\bibfnamefont{R.}~\bibnamefont{Shyam}},
  \bibinfo{journal}{Phys. Rev. C} \textbf{\bibinfo{volume}{61}},
  \bibinfo{pages}{047301} (\bibinfo{year}{2000}).

\bibitem[{\citenamefont{Tostevin et~al.}(1998)\citenamefont{Tostevin, Rugmai,
  and Johnson}}]{PhysRevC.57.3225}
\bibinfo{author}{\bibfnamefont{J.~A.} \bibnamefont{Tostevin}},
  \bibinfo{author}{\bibfnamefont{S.}~\bibnamefont{Rugmai}}, \bibnamefont{and}
  \bibinfo{author}{\bibfnamefont{R.~C.} \bibnamefont{Johnson}},
  \bibinfo{journal}{Phys. Rev. C} \textbf{\bibinfo{volume}{57}},
  \bibinfo{pages}{3225} (\bibinfo{year}{1998}).

\bibitem[{\citenamefont{Faddeev}(1960)}]{faddeev60}
\bibinfo{author}{\bibfnamefont{L.~D.} \bibnamefont{Faddeev}},
  \bibinfo{journal}{Zh. Eksp. Theor. Fiz.} \textbf{\bibinfo{volume}{39}},
  \bibinfo{pages}{1459} (\bibinfo{year}{1960}), \bibinfo{note}{{[Sov. Phys.
  JETP {\bf 12}, 1014 (1961)]}}.

\bibitem[{\citenamefont{Alt et~al.}(1967)\citenamefont{Alt, Grassberger, and
  Sandhas}}]{Alt}
\bibinfo{author}{\bibfnamefont{E.~O.} \bibnamefont{Alt}},
  \bibinfo{author}{\bibfnamefont{P.}~\bibnamefont{Grassberger}},
  \bibnamefont{and} \bibinfo{author}{\bibfnamefont{W.}~\bibnamefont{Sandhas}},
  \bibinfo{journal}{Nucl. Phys. B} \textbf{\bibinfo{volume}{2}},
  \bibinfo{pages}{167} (\bibinfo{year}{1967}).

\bibitem[{\citenamefont{Typel and Baur}(1994)}]{PhysRevC.50.2104}
\bibinfo{author}{\bibfnamefont{S.}~\bibnamefont{Typel}} \bibnamefont{and}
  \bibinfo{author}{\bibfnamefont{G.}~\bibnamefont{Baur}},
  \bibinfo{journal}{Phys. Rev. C} \textbf{\bibinfo{volume}{50}},
  \bibinfo{pages}{2104} (\bibinfo{year}{1994}).

\bibitem[{\citenamefont{Esbensen and Bertsch}(1996)}]{Esbensen199637}
\bibinfo{author}{\bibfnamefont{H.}~\bibnamefont{Esbensen}} \bibnamefont{and}
  \bibinfo{author}{\bibfnamefont{G.}~\bibnamefont{Bertsch}},
  \bibinfo{journal}{Nuclear Physics A} \textbf{\bibinfo{volume}{600}},
  \bibinfo{pages}{37 } (\bibinfo{year}{1996}), ISSN \bibinfo{issn}{0375-9474}.

\bibitem[{\citenamefont{Kido et~al.}(1994)\citenamefont{Kido, Yabana, and
  Suzuki}}]{PhysRevC.50.R1276}
\bibinfo{author}{\bibfnamefont{T.}~\bibnamefont{Kido}},
  \bibinfo{author}{\bibfnamefont{K.}~\bibnamefont{Yabana}}, \bibnamefont{and}
  \bibinfo{author}{\bibfnamefont{Y.}~\bibnamefont{Suzuki}},
  \bibinfo{journal}{Phys. Rev. C} \textbf{\bibinfo{volume}{50}},
  \bibinfo{pages}{R1276} (\bibinfo{year}{1994}).

\bibitem[{\citenamefont{Typel and Baur}(2001)}]{PhysRevC.64.024601}
\bibinfo{author}{\bibfnamefont{S.}~\bibnamefont{Typel}} \bibnamefont{and}
  \bibinfo{author}{\bibfnamefont{G.}~\bibnamefont{Baur}},
  \bibinfo{journal}{Phys. Rev. C} \textbf{\bibinfo{volume}{64}},
  \bibinfo{pages}{024601} (\bibinfo{year}{2001}).

\bibitem[{\citenamefont{Capel et~al.}(2004)\citenamefont{Capel, Goldstein, and
  Baye}}]{PhysRevC.70.064605}
\bibinfo{author}{\bibfnamefont{P.}~\bibnamefont{Capel}},
  \bibinfo{author}{\bibfnamefont{G.}~\bibnamefont{Goldstein}},
  \bibnamefont{and} \bibinfo{author}{\bibfnamefont{D.}~\bibnamefont{Baye}},
  \bibinfo{journal}{Phys. Rev. C} \textbf{\bibinfo{volume}{70}},
  \bibinfo{pages}{064605} (\bibinfo{year}{2004}).

\bibitem[{\citenamefont{Garc\'ia-Camacho
  et~al.}(2006)\citenamefont{Garc\'ia-Camacho, Bonaccorso, and
  Brink}}]{GarcíaCamacho2006118}
\bibinfo{author}{\bibfnamefont{A.}~\bibnamefont{Garc\'ia-Camacho}},
  \bibinfo{author}{\bibfnamefont{A.}~\bibnamefont{Bonaccorso}},
  \bibnamefont{and} \bibinfo{author}{\bibfnamefont{D.}~\bibnamefont{Brink}},
  \bibinfo{journal}{Nuclear Physics A} \textbf{\bibinfo{volume}{776}},
  \bibinfo{pages}{118 } (\bibinfo{year}{2006}), ISSN \bibinfo{issn}{0375-9474}.

\bibitem[{\citenamefont{Summers et~al.}(2006)\citenamefont{Summers, Nunes, and
  Thompson}}]{summers}
\bibinfo{author}{\bibfnamefont{N.~C.} \bibnamefont{Summers}},
  \bibinfo{author}{\bibfnamefont{F.~M.} \bibnamefont{Nunes}}, \bibnamefont{and}
  \bibinfo{author}{\bibfnamefont{I.~J.} \bibnamefont{Thompson}},
  \bibinfo{journal}{Phys. Rev. C} \textbf{\bibinfo{volume}{74}},
  \bibinfo{pages}{014606} (\bibinfo{year}{2006}).

\bibitem[{\citenamefont{de~Diego et~al.}(2014)\citenamefont{de~Diego, Arias,
  Lay, and Moro}}]{deDiego14}
\bibinfo{author}{\bibfnamefont{R.}~\bibnamefont{de~Diego}},
  \bibinfo{author}{\bibfnamefont{J.~M.} \bibnamefont{Arias}},
  \bibinfo{author}{\bibfnamefont{J.~A.} \bibnamefont{Lay}}, \bibnamefont{and}
  \bibinfo{author}{\bibfnamefont{A.~M.} \bibnamefont{Moro}},
  \bibinfo{journal}{Phys. Rev. C} \textbf{\bibinfo{volume}{89}},
  \bibinfo{pages}{064609} (\bibinfo{year}{2014}).

\bibitem[{\citenamefont{Deltuva}(2013)}]{Del13}
\bibinfo{author}{\bibfnamefont{A.}~\bibnamefont{Deltuva}},
  \bibinfo{journal}{Phys. Rev. C} \textbf{\bibinfo{volume}{88}},
  \bibinfo{pages}{011601} (\bibinfo{year}{2013}).

\bibitem[{\citenamefont{Moro and Lay}(2012)}]{Mor12b}
\bibinfo{author}{\bibfnamefont{A.~M.} \bibnamefont{Moro}} \bibnamefont{and}
  \bibinfo{author}{\bibfnamefont{J.~A.} \bibnamefont{Lay}},
  \bibinfo{journal}{Phys. Rev. Lett.} \textbf{\bibinfo{volume}{109}},
  \bibinfo{pages}{232502} (\bibinfo{year}{2012}).

\bibitem[{\citenamefont{Lay et~al.}(2016)\citenamefont{Lay, de~Diego, Crespo,
  Moro, Arias, and Johnson}}]{PhysRevC.94.021602}
\bibinfo{author}{\bibfnamefont{J.~A.} \bibnamefont{Lay}},
  \bibinfo{author}{\bibfnamefont{R.}~\bibnamefont{de~Diego}},
  \bibinfo{author}{\bibfnamefont{R.}~\bibnamefont{Crespo}},
  \bibinfo{author}{\bibfnamefont{A.~M.} \bibnamefont{Moro}},
  \bibinfo{author}{\bibfnamefont{J.~M.} \bibnamefont{Arias}}, \bibnamefont{and}
  \bibinfo{author}{\bibfnamefont{R.~C.} \bibnamefont{Johnson}},
  \bibinfo{journal}{Phys. Rev. C} \textbf{\bibinfo{volume}{94}},
  \bibinfo{pages}{021602} (\bibinfo{year}{2016}).

\bibitem[{\citenamefont{Kiss et~al.}(1976)\citenamefont{Kiss, Aspelund,
  Hrehuss, Knöpfle, Rogge, Schwinn, Seres, Turek, and
  Mayer-Böricke}}]{Kiss19761}
\bibinfo{author}{\bibfnamefont{A.}~\bibnamefont{Kiss}},
  \bibinfo{author}{\bibfnamefont{O.}~\bibnamefont{Aspelund}},
  \bibinfo{author}{\bibfnamefont{G.}~\bibnamefont{Hrehuss}},
  \bibinfo{author}{\bibfnamefont{K.}~\bibnamefont{Knöpfle}},
  \bibinfo{author}{\bibfnamefont{M.}~\bibnamefont{Rogge}},
  \bibinfo{author}{\bibfnamefont{U.}~\bibnamefont{Schwinn}},
  \bibinfo{author}{\bibfnamefont{Z.}~\bibnamefont{Seres}},
  \bibinfo{author}{\bibfnamefont{P.}~\bibnamefont{Turek}}, \bibnamefont{and}
  \bibinfo{author}{\bibfnamefont{C.}~\bibnamefont{Mayer-Böricke}},
  \bibinfo{journal}{Nuclear Physics A} \textbf{\bibinfo{volume}{262}},
  \bibinfo{pages}{1 } (\bibinfo{year}{1976}), ISSN \bibinfo{issn}{0375-9474}.

\bibitem[{\citenamefont{Woodard et~al.}(2012)\citenamefont{Woodard, Figueira,
  Otomar, Niello, Lubian, Arazi, Capurro, Carnelli, Fimiani, Martí
  et~al.}}]{Woodard201217}
\bibinfo{author}{\bibfnamefont{A.}~\bibnamefont{Woodard}},
  \bibinfo{author}{\bibfnamefont{J.}~\bibnamefont{Figueira}},
  \bibinfo{author}{\bibfnamefont{D.}~\bibnamefont{Otomar}},
  \bibinfo{author}{\bibfnamefont{J.~F.} \bibnamefont{Niello}},
  \bibinfo{author}{\bibfnamefont{J.}~\bibnamefont{Lubian}},
  \bibinfo{author}{\bibfnamefont{A.}~\bibnamefont{Arazi}},
  \bibinfo{author}{\bibfnamefont{O.}~\bibnamefont{Capurro}},
  \bibinfo{author}{\bibfnamefont{P.}~\bibnamefont{Carnelli}},
  \bibinfo{author}{\bibfnamefont{L.}~\bibnamefont{Fimiani}},
  \bibinfo{author}{\bibfnamefont{G.}~\bibnamefont{Martí}},
  \bibnamefont{et~al.}, \bibinfo{journal}{Nuclear Physics A}
  \textbf{\bibinfo{volume}{873}}, \bibinfo{pages}{17 } (\bibinfo{year}{2012}),
  ISSN \bibinfo{issn}{0375-9474}.

\bibitem[{\citenamefont{Yahiro et~al.}(1986)\citenamefont{Yahiro, Iseri,
  Kameyama, Kamimura, and Kawai}}]{Yahiro01041986}
\bibinfo{author}{\bibfnamefont{M.}~\bibnamefont{Yahiro}},
  \bibinfo{author}{\bibfnamefont{Y.}~\bibnamefont{Iseri}},
  \bibinfo{author}{\bibfnamefont{H.}~\bibnamefont{Kameyama}},
  \bibinfo{author}{\bibfnamefont{M.}~\bibnamefont{Kamimura}}, \bibnamefont{and}
  \bibinfo{author}{\bibfnamefont{M.}~\bibnamefont{Kawai}},
  \bibinfo{journal}{Progress of Theoretical Physics Supplement}
  \textbf{\bibinfo{volume}{89}}, \bibinfo{pages}{32} (\bibinfo{year}{1986}).

\bibitem[{\citenamefont{Pierre~Chau}(2015)}]{PierreChau2015}
\bibinfo{author}{\bibfnamefont{H.-T.} \bibnamefont{Pierre~Chau}},
  \bibinfo{journal}{The European Physical Journal A}
  \textbf{\bibinfo{volume}{51}}, \bibinfo{pages}{1} (\bibinfo{year}{2015}),
  ISSN \bibinfo{issn}{1434-601X}.

\bibitem[{\citenamefont{Lay et~al.}(2012)\citenamefont{Lay, Moro, Arias, and
  G\'omez-Camacho}}]{particlemotion}
\bibinfo{author}{\bibfnamefont{J.~A.} \bibnamefont{Lay}},
  \bibinfo{author}{\bibfnamefont{A.~M.} \bibnamefont{Moro}},
  \bibinfo{author}{\bibfnamefont{J.~M.} \bibnamefont{Arias}}, \bibnamefont{and}
  \bibinfo{author}{\bibfnamefont{J.}~\bibnamefont{G\'omez-Camacho}},
  \bibinfo{journal}{Phys. Rev. C} \textbf{\bibinfo{volume}{85}},
  \bibinfo{pages}{054618} (\bibinfo{year}{2012}).

\bibitem[{\citenamefont{Thompson}(1988)}]{fresco}
\bibinfo{author}{\bibfnamefont{I.~J.} \bibnamefont{Thompson}},
  \bibinfo{journal}{Computer Physics Reports} \textbf{\bibinfo{volume}{7}},
  \bibinfo{pages}{167 } (\bibinfo{year}{1988}), ISSN \bibinfo{issn}{0167-7977}.

\bibitem[{\citenamefont{Bohr and Mottelson}(1998)}]{bohrmottelson}
\bibinfo{author}{\bibfnamefont{A.}~\bibnamefont{Bohr}} \bibnamefont{and}
  \bibinfo{author}{\bibfnamefont{B.~R.} \bibnamefont{Mottelson}},
  \emph{\bibinfo{title}{Nuclear Structure}} (\bibinfo{publisher}{World
  Scientific Publishing}, \bibinfo{year}{1998}).

\bibitem[{\citenamefont{Moro et~al.}(2009)\citenamefont{Moro, Arias,
  G{\'o}mez-Camacho, and P{\'e}rez-Bernal}}]{moro2009analytical}
\bibinfo{author}{\bibfnamefont{A.}~\bibnamefont{Moro}},
  \bibinfo{author}{\bibfnamefont{J.}~\bibnamefont{Arias}},
  \bibinfo{author}{\bibfnamefont{J.}~\bibnamefont{G{\'o}mez-Camacho}},
  \bibnamefont{and}
  \bibinfo{author}{\bibfnamefont{F.}~\bibnamefont{P{\'e}rez-Bernal}},
  \bibinfo{journal}{Physical Review C} \textbf{\bibinfo{volume}{80}},
  \bibinfo{pages}{054605} (\bibinfo{year}{2009}).

\bibitem[{\citenamefont{Koning and Delaroche}(2003)}]{Koning2003231}
\bibinfo{author}{\bibfnamefont{A.}~\bibnamefont{Koning}} \bibnamefont{and}
  \bibinfo{author}{\bibfnamefont{J.}~\bibnamefont{Delaroche}},
  \bibinfo{journal}{Nuclear Physics A} \textbf{\bibinfo{volume}{713}},
  \bibinfo{pages}{231 } (\bibinfo{year}{2003}), ISSN \bibinfo{issn}{0375-9474}.

\bibitem[{\citenamefont{Sonzogni}(2008)}]{chartnuclides}
\bibinfo{author}{\bibfnamefont{A.}~\bibnamefont{Sonzogni}},
  \bibinfo{journal}{National Nuclear Data Center: Brookhaven National
  Laboratory. Retrieved} \textbf{\bibinfo{volume}{6}} (\bibinfo{year}{2008}).

\bibitem[{\citenamefont{Stephenson}(1980)}]{stephenson1980proceedings}
\bibinfo{author}{\bibfnamefont{E.}~\bibnamefont{Stephenson}},
  \bibinfo{journal}{Proceedings of Fifth International Symposium on
  Polarization Phenomena in Nuclear Physics, Santa Fe}  (\bibinfo{year}{1980}).

\bibitem[{\citenamefont{Deltuva}(2016)}]{Deltuva2016173}
\bibinfo{author}{\bibfnamefont{A.}~\bibnamefont{Deltuva}},
  \bibinfo{journal}{Nuclear Physics A} \textbf{\bibinfo{volume}{947}},
  \bibinfo{pages}{173 } (\bibinfo{year}{2016}), ISSN \bibinfo{issn}{0375-9474}.

\bibitem[{\citenamefont{Varner et~al.}(1991)\citenamefont{Varner, Thompson,
  McAbee, Ludwig, and Clegg}}]{CH89}
\bibinfo{author}{\bibfnamefont{R.}~\bibnamefont{Varner}},
  \bibinfo{author}{\bibfnamefont{W.}~\bibnamefont{Thompson}},
  \bibinfo{author}{\bibfnamefont{T.}~\bibnamefont{McAbee}},
  \bibinfo{author}{\bibfnamefont{E.}~\bibnamefont{Ludwig}}, \bibnamefont{and}
  \bibinfo{author}{\bibfnamefont{T.}~\bibnamefont{Clegg}},
  \bibinfo{journal}{Physics Reports} \textbf{\bibinfo{volume}{201}},
  \bibinfo{pages}{57 } (\bibinfo{year}{1991}), ISSN \bibinfo{issn}{0370-1573}.

\bibitem[{\citenamefont{Machleidt}(2001)}]{Bonn}
\bibinfo{author}{\bibfnamefont{R.}~\bibnamefont{Machleidt}},
  \bibinfo{journal}{Phys. Rev. C} \textbf{\bibinfo{volume}{63}},
  \bibinfo{pages}{024001} (\bibinfo{year}{2001}).

\bibitem[{\citenamefont{De~Leo et~al.}(1981)\citenamefont{De~Leo, D'Erasmo,
  Pantaleo, Harakeh, Micheletti, and Pignanelli}}]{PhysRevC.23.1355}
\bibinfo{author}{\bibfnamefont{R.}~\bibnamefont{De~Leo}},
  \bibinfo{author}{\bibfnamefont{G.}~\bibnamefont{D'Erasmo}},
  \bibinfo{author}{\bibfnamefont{A.}~\bibnamefont{Pantaleo}},
  \bibinfo{author}{\bibfnamefont{M.~N.} \bibnamefont{Harakeh}},
  \bibinfo{author}{\bibfnamefont{S.}~\bibnamefont{Micheletti}},
  \bibnamefont{and}
  \bibinfo{author}{\bibfnamefont{M.}~\bibnamefont{Pignanelli}},
  \bibinfo{journal}{Phys. Rev. C} \textbf{\bibinfo{volume}{23}},
  \bibinfo{pages}{1355} (\bibinfo{year}{1981}).

\bibitem[{\citenamefont{Thompson and Nunes}(2009)}]{thompsonnunes}
\bibinfo{author}{\bibfnamefont{I.~J.} \bibnamefont{Thompson}} \bibnamefont{and}
  \bibinfo{author}{\bibfnamefont{F.~M.} \bibnamefont{Nunes}},
  \emph{\bibinfo{title}{Nuclear Reactions for Astrophysics}}
  (\bibinfo{publisher}{Cambridge University Press}, \bibinfo{year}{2009}).

\bibitem[{\citenamefont{RAMAN et~al.}(2001)\citenamefont{RAMAN, JR., and
  TIKKANEN}}]{RAMAN20011}
\bibinfo{author}{\bibfnamefont{S.}~\bibnamefont{RAMAN}},
  \bibinfo{author}{\bibfnamefont{C.~N.} \bibnamefont{JR.}}, \bibnamefont{and}
  \bibinfo{author}{\bibfnamefont{P.}~\bibnamefont{TIKKANEN}},
  \bibinfo{journal}{Atomic Data and Nuclear Data Tables}
  \textbf{\bibinfo{volume}{78}}, \bibinfo{pages}{1 } (\bibinfo{year}{2001}),
  ISSN \bibinfo{issn}{0092-640X}.

\bibitem[{\citenamefont{KIBÉDI and SPEAR}(2002)}]{KIBEDI200235}
\bibinfo{author}{\bibfnamefont{T.}~\bibnamefont{KIBÉDI}} \bibnamefont{and}
  \bibinfo{author}{\bibfnamefont{R.}~\bibnamefont{SPEAR}},
  \bibinfo{journal}{Atomic Data and Nuclear Data Tables}
  \textbf{\bibinfo{volume}{80}}, \bibinfo{pages}{35 } (\bibinfo{year}{2002}),
  ISSN \bibinfo{issn}{0092-640X}.

\bibitem[{\citenamefont{Karataglidis et~al.}(2005)\citenamefont{Karataglidis,
  Amos, and Giraud}}]{PhysRevC.71.064601}
\bibinfo{author}{\bibfnamefont{S.}~\bibnamefont{Karataglidis}},
  \bibinfo{author}{\bibfnamefont{K.}~\bibnamefont{Amos}}, \bibnamefont{and}
  \bibinfo{author}{\bibfnamefont{B.~G.} \bibnamefont{Giraud}},
  \bibinfo{journal}{Phys. Rev. C} \textbf{\bibinfo{volume}{71}},
  \bibinfo{pages}{064601} (\bibinfo{year}{2005}).

\bibitem[{\citenamefont{Matsumoto et~al.}(2003)\citenamefont{Matsumoto,
  Kamizato, Ogata, Iseri, Hiyama, Kamimura, and Yahiro}}]{PhysRevC.68.064607}
\bibinfo{author}{\bibfnamefont{T.}~\bibnamefont{Matsumoto}},
  \bibinfo{author}{\bibfnamefont{T.}~\bibnamefont{Kamizato}},
  \bibinfo{author}{\bibfnamefont{K.}~\bibnamefont{Ogata}},
  \bibinfo{author}{\bibfnamefont{Y.}~\bibnamefont{Iseri}},
  \bibinfo{author}{\bibfnamefont{E.}~\bibnamefont{Hiyama}},
  \bibinfo{author}{\bibfnamefont{M.}~\bibnamefont{Kamimura}}, \bibnamefont{and}
  \bibinfo{author}{\bibfnamefont{M.}~\bibnamefont{Yahiro}},
  \bibinfo{journal}{Phys. Rev. C} \textbf{\bibinfo{volume}{68}},
  \bibinfo{pages}{064607} (\bibinfo{year}{2003}).

\bibitem[{\citenamefont{Chamon et~al.}(1997)\citenamefont{Chamon, Pereira,
  Hussein, C\^andido~Ribeiro, and Galetti}}]{PhysRevLett.79.5218}
\bibinfo{author}{\bibfnamefont{L.~C.} \bibnamefont{Chamon}},
  \bibinfo{author}{\bibfnamefont{D.}~\bibnamefont{Pereira}},
  \bibinfo{author}{\bibfnamefont{M.~S.} \bibnamefont{Hussein}},
  \bibinfo{author}{\bibfnamefont{M.~A.} \bibnamefont{C\^andido~Ribeiro}},
  \bibnamefont{and} \bibinfo{author}{\bibfnamefont{D.}~\bibnamefont{Galetti}},
  \bibinfo{journal}{Phys. Rev. Lett.} \textbf{\bibinfo{volume}{79}},
  \bibinfo{pages}{5218} (\bibinfo{year}{1997}).

\bibitem[{\citenamefont{Perey and Perey}(1963)}]{PhysRev.132.755}
\bibinfo{author}{\bibfnamefont{C.~M.} \bibnamefont{Perey}} \bibnamefont{and}
  \bibinfo{author}{\bibfnamefont{F.~G.} \bibnamefont{Perey}},
  \bibinfo{journal}{Phys. Rev.} \textbf{\bibinfo{volume}{132}},
  \bibinfo{pages}{755} (\bibinfo{year}{1963}).

\bibitem[{\citenamefont{Hirabayashi and Sakuragi}(1991)}]{Hirabayashi199111}
\bibinfo{author}{\bibfnamefont{Y.}~\bibnamefont{Hirabayashi}} \bibnamefont{and}
  \bibinfo{author}{\bibfnamefont{Y.}~\bibnamefont{Sakuragi}},
  \bibinfo{journal}{Physics Letters B} \textbf{\bibinfo{volume}{258}},
  \bibinfo{pages}{11 } (\bibinfo{year}{1991}), ISSN \bibinfo{issn}{0370-2693}.

\bibitem[{\citenamefont{Santra et~al.}(2009)\citenamefont{Santra, Parkar,
  Ramachandran, Pal, Shrivastava, Roy, Nayak, Chatterjee, Choudhury, and
  Kailas}}]{Santra2009139}
\bibinfo{author}{\bibfnamefont{S.}~\bibnamefont{Santra}},
  \bibinfo{author}{\bibfnamefont{V.}~\bibnamefont{Parkar}},
  \bibinfo{author}{\bibfnamefont{K.}~\bibnamefont{Ramachandran}},
  \bibinfo{author}{\bibfnamefont{U.}~\bibnamefont{Pal}},
  \bibinfo{author}{\bibfnamefont{A.}~\bibnamefont{Shrivastava}},
  \bibinfo{author}{\bibfnamefont{B.}~\bibnamefont{Roy}},
  \bibinfo{author}{\bibfnamefont{B.}~\bibnamefont{Nayak}},
  \bibinfo{author}{\bibfnamefont{A.}~\bibnamefont{Chatterjee}},
  \bibinfo{author}{\bibfnamefont{R.}~\bibnamefont{Choudhury}},
  \bibnamefont{and} \bibinfo{author}{\bibfnamefont{S.}~\bibnamefont{Kailas}},
  \bibinfo{journal}{Physics Letters B} \textbf{\bibinfo{volume}{677}},
  \bibinfo{pages}{139 } (\bibinfo{year}{2009}), ISSN \bibinfo{issn}{0370-2693}.

\bibitem[{\citenamefont{Beck et~al.}(2007)\citenamefont{Beck, Keeley, and
  Diaz-Torres}}]{PhysRevC.75.054605}
\bibinfo{author}{\bibfnamefont{C.}~\bibnamefont{Beck}},
  \bibinfo{author}{\bibfnamefont{N.}~\bibnamefont{Keeley}}, \bibnamefont{and}
  \bibinfo{author}{\bibfnamefont{A.}~\bibnamefont{Diaz-Torres}},
  \bibinfo{journal}{Phys. Rev. C} \textbf{\bibinfo{volume}{75}},
  \bibinfo{pages}{054605} (\bibinfo{year}{2007}).

\bibitem[{\citenamefont{Watanabe et~al.}(2012)\citenamefont{Watanabe,
  Matsumoto, Minomo, Ogata, and Yahiro}}]{PhysRevC.86.031601}
\bibinfo{author}{\bibfnamefont{S.}~\bibnamefont{Watanabe}},
  \bibinfo{author}{\bibfnamefont{T.}~\bibnamefont{Matsumoto}},
  \bibinfo{author}{\bibfnamefont{K.}~\bibnamefont{Minomo}},
  \bibinfo{author}{\bibfnamefont{K.}~\bibnamefont{Ogata}}, \bibnamefont{and}
  \bibinfo{author}{\bibfnamefont{M.}~\bibnamefont{Yahiro}},
  \bibinfo{journal}{Phys. Rev. C} \textbf{\bibinfo{volume}{86}},
  \bibinfo{pages}{031601} (\bibinfo{year}{2012}).

\bibitem[{\citenamefont{Cook}(1982)}]{Cook1982153}
\bibinfo{author}{\bibfnamefont{J.}~\bibnamefont{Cook}},
  \bibinfo{journal}{Nuclear Physics A} \textbf{\bibinfo{volume}{388}},
  \bibinfo{pages}{153 } (\bibinfo{year}{1982}), ISSN \bibinfo{issn}{0375-9474}.

\bibitem[{\citenamefont{Muskat et~al.}(1995)\citenamefont{Muskat, Carter,
  Fearick, and Hnizdo}}]{MUSKAT199542}
\bibinfo{author}{\bibfnamefont{E.}~\bibnamefont{Muskat}},
  \bibinfo{author}{\bibfnamefont{J.}~\bibnamefont{Carter}},
  \bibinfo{author}{\bibfnamefont{R.}~\bibnamefont{Fearick}}, \bibnamefont{and}
  \bibinfo{author}{\bibfnamefont{V.}~\bibnamefont{Hnizdo}},
  \bibinfo{journal}{Nuclear Physics A} \textbf{\bibinfo{volume}{581}},
  \bibinfo{pages}{42 } (\bibinfo{year}{1995}), ISSN \bibinfo{issn}{0375-9474}.

\bibitem[{\citenamefont{Diaz-Torres et~al.}(2003)\citenamefont{Diaz-Torres,
  Thompson, and Beck}}]{PhysRevC.68.044607}
\bibinfo{author}{\bibfnamefont{A.}~\bibnamefont{Diaz-Torres}},
  \bibinfo{author}{\bibfnamefont{I.~J.} \bibnamefont{Thompson}},
  \bibnamefont{and} \bibinfo{author}{\bibfnamefont{C.}~\bibnamefont{Beck}},
  \bibinfo{journal}{Phys. Rev. C} \textbf{\bibinfo{volume}{68}},
  \bibinfo{pages}{044607} (\bibinfo{year}{2003}).

\bibitem[{\citenamefont{Taylor et~al.}(1965)\citenamefont{Taylor, Fletcher, and
  Davis}}]{Taylor1965318}
\bibinfo{author}{\bibfnamefont{R.}~\bibnamefont{Taylor}},
  \bibinfo{author}{\bibfnamefont{N.}~\bibnamefont{Fletcher}}, \bibnamefont{and}
  \bibinfo{author}{\bibfnamefont{R.}~\bibnamefont{Davis}},
  \bibinfo{journal}{Nuclear Physics} \textbf{\bibinfo{volume}{65}},
  \bibinfo{pages}{318 } (\bibinfo{year}{1965}), ISSN \bibinfo{issn}{0029-5582}.

\bibitem[{\citenamefont{Pires et~al.}(2011)\citenamefont{Pires,
  Lichtenth\"aler, L\'epine-Szily, Guimar\~aes, de~Faria, Barioni,
  Mendes~Junior, Morcelle, Pampa~Condori, Morais et~al.}}]{PhysRevC.83.064603}
\bibinfo{author}{\bibfnamefont{K.~C.~C.} \bibnamefont{Pires}},
  \bibinfo{author}{\bibfnamefont{R.}~\bibnamefont{Lichtenth\"aler}},
  \bibinfo{author}{\bibfnamefont{A.}~\bibnamefont{L\'epine-Szily}},
  \bibinfo{author}{\bibfnamefont{V.}~\bibnamefont{Guimar\~aes}},
  \bibinfo{author}{\bibfnamefont{P.~N.} \bibnamefont{de~Faria}},
  \bibinfo{author}{\bibfnamefont{A.}~\bibnamefont{Barioni}},
  \bibinfo{author}{\bibfnamefont{D.~R.} \bibnamefont{Mendes~Junior}},
  \bibinfo{author}{\bibfnamefont{V.}~\bibnamefont{Morcelle}},
  \bibinfo{author}{\bibfnamefont{R.}~\bibnamefont{Pampa~Condori}},
  \bibinfo{author}{\bibfnamefont{M.~C.} \bibnamefont{Morais}},
  \bibnamefont{et~al.}, \bibinfo{journal}{Phys. Rev. C}
  \textbf{\bibinfo{volume}{83}}, \bibinfo{pages}{064603}
  (\bibinfo{year}{2011}).

\bibitem[{\citenamefont{Dave and Gould}(1983)}]{PhysRevC.28.2212}
\bibinfo{author}{\bibfnamefont{J.~H.} \bibnamefont{Dave}} \bibnamefont{and}
  \bibinfo{author}{\bibfnamefont{C.~R.} \bibnamefont{Gould}},
  \bibinfo{journal}{Phys. Rev. C} \textbf{\bibinfo{volume}{28}},
  \bibinfo{pages}{2212} (\bibinfo{year}{1983}).

\bibitem[{\citenamefont{Bonaccorso and Charity}(2014)}]{PhysRevC.89.024619}
\bibinfo{author}{\bibfnamefont{A.}~\bibnamefont{Bonaccorso}} \bibnamefont{and}
  \bibinfo{author}{\bibfnamefont{R.~J.} \bibnamefont{Charity}},
  \bibinfo{journal}{Phys. Rev. C} \textbf{\bibinfo{volume}{89}},
  \bibinfo{pages}{024619} (\bibinfo{year}{2014}).

\end{thebibliography}

\end{document}